\documentclass{sig-alternate-05-2015}

\usepackage{tikz}
\usetikzlibrary{calc}
\usepackage{url}
\usepackage{mdframed}
\usepackage{algpseudocode}
\usepackage{pgfplots}
\usepackage{pgfplotstable}
\usepackage{graphicx}
\usepackage{subcaption}
\usepackage{caption}
\usetikzlibrary{calc}
\usetikzlibrary{patterns}
\usetikzlibrary{shapes.multipart, arrows}
\usetikzlibrary{decorations.pathreplacing}
\usepackage{multirow}

\pgfplotsset{compat=newest}

\def\pdfrate{$\textproc{PDFrate}$}

\def\evadehc{{\tt EvadeHC}}
\def\evader{{\tt BiRand}}
\def\exhaust{{\tt SeqRand}}
\def\evadegp{{\tt EvadeGP}}

\definecolor{cornellred}{rgb}{0.7, 0.11, 0.11}
\definecolor{arsenic}{rgb}{0.23, 0.27, 0.29}
\definecolor{auburn}{rgb}{0.43, 0.21, 0.1}
\definecolor{charcoal}{rgb}{0.21, 0.27, 0.31}
\definecolor{deepcarrotorange}{rgb}{0.91, 0.41, 0.17}
\definecolor{eggplant}{rgb}{0.38, 0.25, 0.32}

\makeatletter
\DeclareRobustCommand*\cal{\@fontswitch\relax\mathcal}
\makeatother

\makeatletter
\def\@copyrightspace{\relax}
\makeatother

\begin{document} \sloppy
\title{Evading Classifiers by Morphing in the Dark}
\author{{Hung Dang, Yue Huang, Ee-Chien Chang}\\
School of Computing, National University of Singapore\\
}

\maketitle

\section*{Abstract}

Learning-based systems have been shown to be vulnerable to adversarial data manipulation attacks. These attacks have been studied under assumptions that the adversary has certain knowledge of either the target model internals, its training dataset or at least classification scores it assigns to input samples. In this paper, we investigate a much more constrained and realistic attack scenario that does not assume any of the above mentioned knowledge. The target classifier is minimally exposed to the adversary, revealing on its final classification decision (e.g., reject or accept an input sample). Moreover, the adversary can only manipulate malicious samples using a blackbox morpher. That is, the adversary has to evade the target classifier by morphing malicious samples ``in the dark''. We present a scoring mechanism that can assign a real-value score which reflects evasion progress to each sample based on limited information available. Leveraging on such scoring mechanism, we propose a hill-climbing method, dubbed  \evadehc, that operates without the help of any domain-specific knowledge, and evaluate it against two PDF malware detectors, namely \pdfrate~and Hidost. The experimental evaluation demonstrates that the proposed evasion attacks are effective, attaining $100\%$ evasion rate on our dataset. Interestingly, \evadehc~outperforms the known classifier evasion technique that operates based on classification scores output by the classifiers. Although our evaluations are conducted on PDF malware classifier, the proposed approaches are domain-agnostic and is of wider application to  other learning-based systems.

\section{Introduction}
\label{sec:intro}
Machine learning techniques have witnessed a steady adoption in a wide range of application domains such as image classification~\cite{ML_face} or natural language processing~\cite{ML_grammar}. Various learning-based systems have been reported to achieve high accuracy, even surpassing human-level performance~\cite{ML_image, ML_Go}. Given the assumption that the training datasets are representative, these systems are expected to perform well in reality on operational data. 

Learning methods have also lent itself to security tasks. Numerous innovative applications of machine learning in security contexts, especially for detection of security violations, have been discussed in the literature~\cite{ML_NID, ML_spam_detection, ML_malware_detection}. 
However, when learning-based systems are deployed for security applications, their accuracy may be challenged by intentional noise and deviations. Various studies \cite{ML_security_conceptual, ML_security_methodical, ML_security_practical} suggest that they are likely susceptible to adversarial data manipulation, for the assumption on representativeness of the training datasets  no longer holds true in the presence of malicious adversaries. For example, a motivated attacker may be able  to morph a malicious instance so that it resembles a benign instance found in the training datasets, evading the existing learning-based classifier.

Several works have studied evasion attacks against learning-based systems, making different assumptions on the amount of knowledge the adversaries have about the system they are attacking. For examples, {\v{S}}rndic et al.~\cite{evade_adhoc_14} and Sharif et al.\cite{face_reg_attack} studied attack scenarios wherein the adversaries have a high level of knowledge about the internals of the target system (e.g., features space and classification algorithm). Xu et al.~\cite{evademl} investigated a more constrained evasion scenario in which the adversaries only have black-box accesses to the target detector that outputs a real-value classification score for an input sample\footnote{Although the approach treats the detector as a blacbox, it still makes use of some feature property. Specifically, the approach exploited a property that a sequence of morphing steps which work for one malicious sample is likely to work for others to accelerate the evasion.}. While the evasion attacks are effective in the presence of the required auxiliary information, the authors of~\cite{evademl} suggested that a  simple  preventive measure of hiding  the classification score would be sufficient to thwart their attacks. Thus, it remains a question how these attacks operate against blackbox systems deployed in reality (e.g., malware detector built-in to email services), for they are unlikely to reveal real-value scores, but rather expose only the final decisions (e.g., reject or accept a sample).

In this work, we study the problem of classifier evasion in the dark, investigating an evasion scenario that is much more constrained and realistic than the ones considered in existing works~\cite{evademl, evade_adhoc_14, membership_infer}. In particular, we consider a very restricted setting in which the adversary does not have any knowledge about the target system (e.g., model internals and training dataset). Moreover, the target system only reveals its final decision instead of a real-value score that reflects its internal state. We further assume that the only feasible mean to manipulate the samples is through some given tools that perform ``random''  morphing. 

To this end, we formulate a model that captures the above-mentioned  restricted setting. Under this model, the adversary is confined to three blackboxes, a binary-output detector (or classifier) that the adversary attempts to evade, a tester that checks whether a sample possesses malicious functionality, and a morpher that transforms the samples. The adversary's goal is to, given a malicious sample, adaptively queries the blackboxes to search for a morphed sample that evades  detection, and yet  maintains its malicious functionality.

We show that under this constrained setting with only blackbox accesses, learning-based systems are still susceptible to evasion attacks. To demonstrate feasibility of effective evasion attacks, we present an evasion method, dubbed \evadehc, which is based on hill-climbing techniques. The main component of \evadehc~is a scoring mechanism that assigns real-value score to the sample based on the binary outcomes obtained from the tester and detector.  The intuition is  to measure the number of morphing steps required to change the detector and tester's decisions, and derive  the score based on these values.  We believe that this scoring mechanism can be used to relax the assumption on the availability of real-value scores that other settings~\cite{watermarking_sensitivity, membership_infer, evademl} make.

We evaluate the proposed evasion technique on two well-known PDF malware classifiers, namely \pdfrate~\cite{pdfrate} and Hidost~\cite{hidost}. In order to enable a fair basis for benchmarking with previous work, we adopt the dataset that is used by Xu et al.~\cite{evademl} in our experiments. This dataset consists of $500$ malicious samples chosen from the Contagio archive~\cite{contagio}. We first compare \evadehc~against a baseline solution which keeps generating random morphed samples and checking them against the tester and detector until an evading sample is found. Empirical results show that \evadehc~attains $100\%$ evasion rate on the experimental dataset, and outperforms the baseline solution by upto $80$ times in term of execution cost. Further, we also experiment on a hypothetical situation wherein the classifiers are hardened by lowering the classification thresholds (i.e., decreasing false acceptance rate at a cost of increasing  false rejection rate), benchmarking \evadehc~against the state-of-the-art method in evading blackbox classifiers~\cite{evademl}. The results strongly demonstrate the robustness of \evadehc. For instance, when the threshold of Hidost is reduced from $0$ to $-0.75$ and the number of detector queries is bounded at $2,500$, \evadehc~attains an evasion rate of as high as $62\%$, in comparison with $8\%$ by the baseline. We also compare \evadehc~with the technique by Xu et al.~\cite{evademl} that uses real-value classification scores during evasion. Interestingly, even with only access to binary outputs of the detector, our approach still outperforms the previous work. While this may appear counter-intuitive, we contend that this result is in fact expected. We believe the reasons are two folds. First, \evadehc~is capable of incorporating information obtained from both the tester and detector, as opposed to previous work relying solely on the classification scores output by the detector. Secondly, the fact that the detector can be evaded implies the classification scores are not a reliable representation of the samples' maliciousness.

\textit{Contributions. } This paper makes the following contributions:
\begin{enumerate}
\item We give a formulation of classifier evasion in the dark  whereby the adversary only has blackbox accesses to the detector,  a morpher and a tester. We also give a probabilisitic model {\tt HsrMopher} to formalise the  notion that no  domain-specific  knowledge can be exploited in the evasion process.

\item We design a scoring function that can assign real-value score reflecting evasion progress to samples, given only binary outcomes obtained from the detector and tester. We believe that this scoring mechanism is useful in extending existing works that necessitate classification scores or other auxiliary information to operate under a more restricted and realistic setting like ours. 

\item Leveraging on the scoring function, we propose an effective hill-climbing based evasion attack {\tt EvadeHC}. This algorithm  is generic in the sense that it does not rely on any domain-specific knowledge of the underlying morphing and detection mechanisms. 

\item We conduct experimental evaluation on two popular PDF malware classifiers. The empirical results demonstrate not only the efficiency but also the robustness of \evadehc. More notably, it is also suggested that the scoring mechanism underlying \evadehc~is more informative than the one that only relies on classification scores~\cite{evademl}.

\end{enumerate}

The rest of the paper is structured as follows. We formulate the problem and discuss its related challenges in Section~\ref{sec:problem} before proposing our evading methods in Section~\ref{sec:proposed_method}. Next, we formalize our proposed approach by presenting probabilistic models in Section~\ref{sec:model} and report the experimental evaluation in Section~\ref{sec:eval}. We discuss the implications of our evasion attacks and their mitigation strategies in Section~\ref{sec:discussion}. We survey related works in Section~\ref{sec:related_work} before concluding our work in Section~\ref{sec:conclusion}.
\newcommand{\morpher}{morpher }

\section{Problem formulation}
\label{sec:problem}
In this section, we define the problem of classifier evasion in the dark and discuss its related challenges. Prior to presenting the formulation, we give a running example to illustrate the problem and its relevant concepts. 

\subsection{Motivating Scenario}

Let us consider an adversary who wants to send a  malware over email channel to a victim. The adversary chooses to embed  the malware in a PDF file, for  the victim  is  more willing to open a PDF file than other file formats. Most email service providers would have built-in malware detection mechanisms scanning users' attachments. Such malware detection mechanism is usually a classifier  that makes decision based on some extracted features, with a classification model trained using existing data. We assume the adversary does not have any knowledge about the detector that the email service provider employs (i.e., the algorithms and feature space adopted by the classifier). Nevertheless, the adversary probes the detector by  sending  emails with the malicious payload  to an account owned by the adversary, and  observing the  binary responses (accept/reject) from the email server. Further, the adversary could adaptively modify his malicious PDF file to search for a PDF file that evades detection.   However,  we assume that the adversary could only probe the email server's detector a limited number of times before it is blacklisted.

The adversary expects that its malicious file will be rejected by the detector. To evade detection, it has access to two tools:  a morphing tool that transforms the PDF sample,  and a sandboxing tool that tests whether the sample maintains its malicious functionalities.  For instance, the morphing tool may insert, delete or replace objects in an underlying representation of the file, and the sandboxing tool dynamically detects whether the malicious PDF sample causes the vulnerable PDF reader to make certain unexpected system calls.  Due to its insufficient understanding of the underlying mechanism to manipulate the PDF sample, the adversary employs the morphing tool as a blackbox. Furthermore, due to the complexity of the PDF  reader,  the adversary does not know whether a morphed PDF sample will retain its  functionality, and the only way to determine that is to invoke the sandbox test.

Given such limitations on the knowledge of both the detector and morphing mechanism, we want to investigate whether effective attacks are still possible. To capture such  constrained capability,  our formulation centres on the notion of  three  blackboxes: a binary-outcome detector $\cal D$ that the adversary wants to evade, a morpher $\cal M$ that  ``randomly'' yet consistently morphs the sample, and a tester $\cal T$ that checks the sample's functionality.

\subsection{Tester $\cal T$, Detector $\cal D$ and Evasion}
The tester $\cal T$, corresponding to the sandbox in the motivation scenario, declares whether a submitted sample $x$ is {\em malicious} or {\em benign}. $\cal T$ is deterministic in the sense that it will output consistent decisions for the same sample. 

The blackbox detector $\cal D$ also takes a sample $x$ as input, and decides whether to {\em accept}
or to {\em reject} the sample.   $\cal D$ corresponds to the malware classifier in the motivating scenario. Samples that are rejected by the detector are those that are classified by the detector to be malicious. It is possible that a sample $x$  declared by $\cal T$ to be malicious is accepted by $\cal D$.  In such case,  we say that the sample $x$ {\em evades}  detection.  Of course, if the detector is exactly the same as the tester, evasion is not possible. In fact, the main objective of our work is to study the security of detectors with imperfect accuracy.  Similar to $\cal T$, we consider detectors that are  deterministic (i.e., their output is always the  same for the same input).

 We highlight that, in our formulation, the detector's output is binary (e.g., accept/reject), as opposed to many previous works (e.g., \cite{evade_adhoc_14,evademl}) which assume real-value outputs.

\subsection{Morpher $\cal M$}
The \morpher~$\cal M$ takes as input a sample $x$ and a random seed $s$,   and deterministically outputs another sample $\widetilde{x}$. We call such action a morphing step. The \morpher~corresponds to the morphing mechanism described in the motivating scenario. The random seed $s$ supplies the randomness required by the morpher.  We are not concerned with the representation of the  random seed $s$, and for simplicity,  treat it as a short binary string.

Starting from a sample $x_0$, the adversary can make successive calls to $\cal M$, say with a sequence of random seeds $ {\bf s} = \langle  s_1,s_2, \ldots,  s_L\rangle$, to obtain a sequence of samples  ${\bf x}=\langle x_0, x_1, x_2, \ldots, x_L \rangle$,  where $x_i = {\cal M}(x_{i-1}, s_i)$.  Let us call $({\bf x}, {\bf s})$ a {\em path} with starting point $x_0$,  endpoint $x_L$, and  path length $L$. When it is clear from the context, we shall omit ${\bf s}$ in the notation, and simply refer to  the path by ${\bf x}$.

The formulation does not dictate how the morphing is to be carried out.  The adversary can exploit useful properties of the application domain to manipulate the samples. For instance, the adversary may be able to efficiently find a morphing path connecting two given samples, or know how  to merge two morphing paths to attain certain desirable property.  Nevertheless, such  domain specific properties are not always available in general.   In Section \ref{sec:model}, we propose a probabilistic model of  ``random'' morphing to capture the restriction that no underlying useful properties can be exploited in manipulating the samples.

\subsection{Adversary's Goal and Performance Cost}

Given a malicious sample $x_0$, the goal of the adversary is to find a sample that evades detection with minimum cost. We call a sample {\em evading} if it is accepted by the detector $\cal D$, but exhibits malicious behaviours as perceived by the tester $\cal T$. If the given sample $x_0$ is already evading, then the adversary has trivially met the goal. Otherwise, the adversary can call the \morpher ${\cal M}$ to obtain other samples, and check the samples by issuing queries to ${\cal T}$ and ${\cal D}$.  

Let $N_{d}, N_{t}$, and $ N_{m}$ be the number of queries the adversary sent to  ${\cal D}$, ${\cal T}$ and ${\cal M}$ over the course of the evasion, respectively.  We are interested in scenarios where $N_d$ is a dominating component in determining the evasion cost. In the motivating scenario, the detector can only be accessed remotely and the email server (who is the defender) imposes a bound on the number of such accesses.  
 
 While  the adversary could freely access the tester, its computational cost might be non-trivial. For instance, in the motivating scenario, computationally intensive dynamic analysis is required to check the functionality. In our experiments, each such test takes  around $45$ seconds on average. Morphing, on the other hands, is less computationally expensive. Hence, it is reasonable to consider an objective function where $N_t$ carries significantly more weight than $N_m$. 
 
A possible optimisation model is to minimise some cost function (e.g., $N_d + 0.5 N_t + 0.01 N_m$).  Alternatively, we could consider constrained optimisation that imposes a bound on  $N_d$, while minimising a cost function involving the other two measurements (e.g., $N_t +  0.02 N_m$).  Since cost functions depend on application scenarios, we do not attempt to tune our algorithm to any particular cost function. Instead, we design search strategies with a general objective of minimising $N_d$, followed by  $N_t$ and $N_m$.

\subsection{Flipping Samples and Gap}
\begin{figure}

\centering
\includegraphics[height=2.3cm]{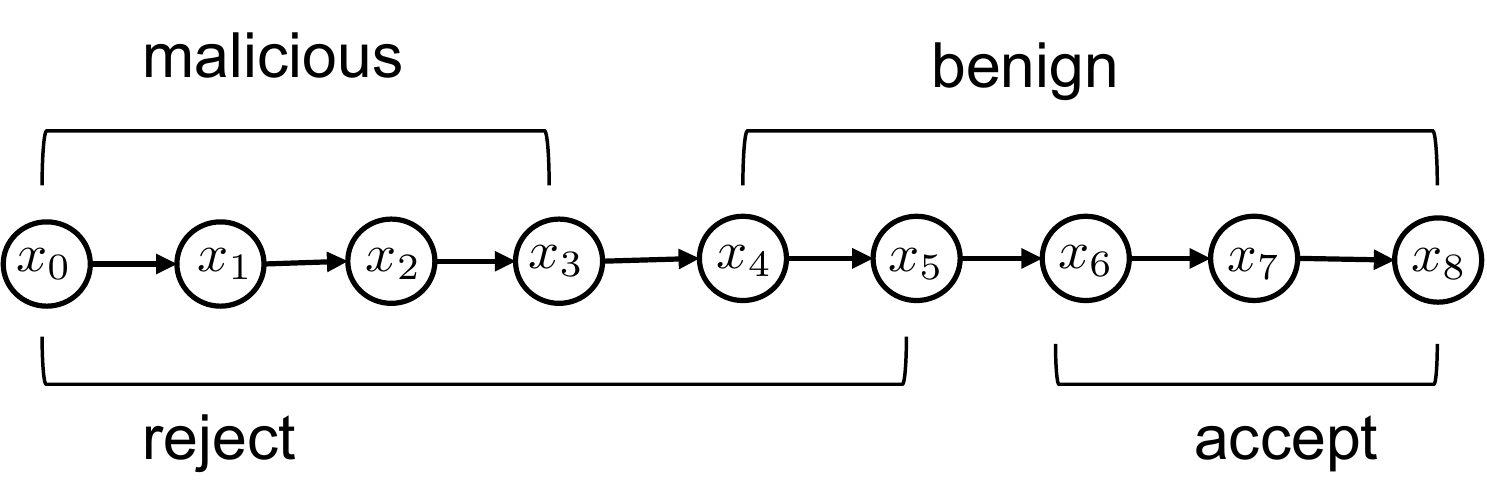} \\
(a)\\
\includegraphics[height=2.3cm]{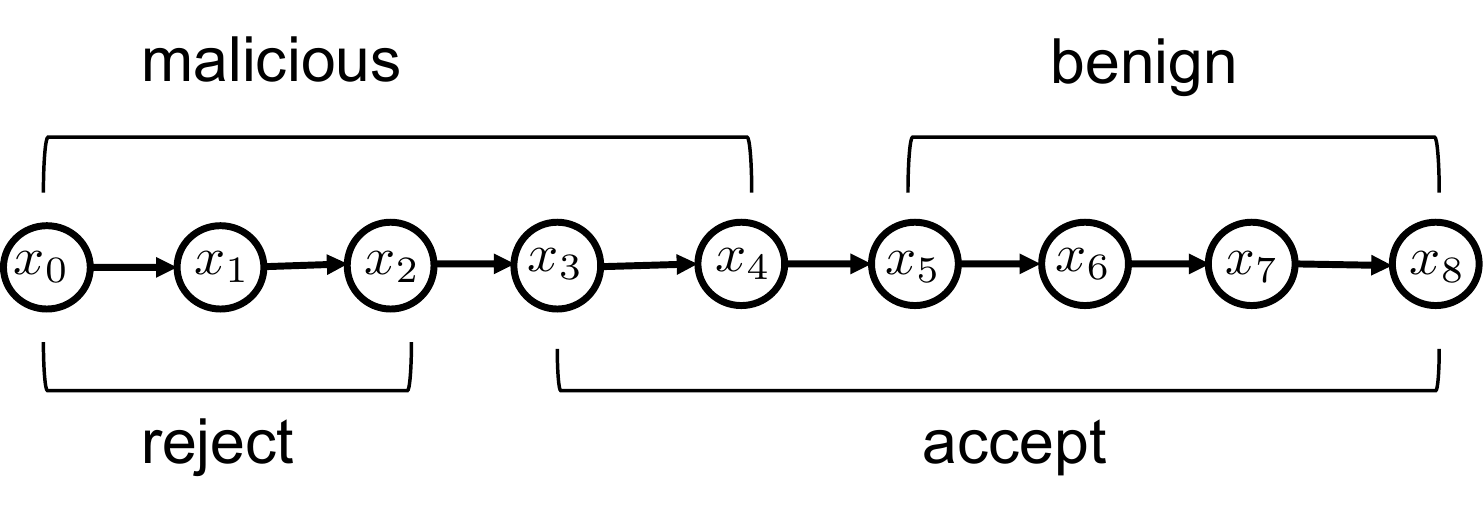} \\
(b)\\
\caption{Illustration of flipping samples.  (a)  The malice-flipping sample is $x_4$ and the reject-flipping sample is $x_6$. The malice-flipping distance   $m_{{\bf x}} = 4$  and  the reject-flipping distance  $r_{\bf x} = 6 $.   (b)  The malice-flipping sample is $x_5$ whereas the reject-flipping sample is $x_3$. The malice-flipping distance   $m_{{\bf x}} = 5$  and  the reject-flipping distance  $r_{\bf x} = 3 $. Since the gap  ($r_{\bf x} - m_{\bf x}$) is  negative,  $x_3$ is evading.   }

\label{fig:flipping}
\end{figure}

Let us consider a typical path ${\bf x}= \langle x_0, x_1, \ldots, x_L\rangle$ in which $x_0$ is malicious and rejected while $x_L$ is benign and accepted.
Somewhere along the path, the samples turn from malicious to benign, and from  rejected to  accepted. Starting from $x_0$, let us call  the first sample that is benign the {\em malice-flipping sample}, and the first accepted sample the  {\em reject-flipping sample} (see Figure \ref{fig:flipping}).    Let $m_{\bf x}$ and $r_{\bf x}$ be the respective  number of morphing steps from $x_0$ to reach the malice-flipping sample and reject-flipping sample. That is, 
  \begin{eqnarray*}
  m _{\bf x} & = & \arg\min_i  \{ {\cal T} (x_i) = {\mbox{\em benign} }  \},  \\
          r_{\bf x} & =&  \arg\min_i   \{ {\cal D} (x_i) = \mbox{\em accept}  \} 
  \end{eqnarray*}       
We call $m_{\bf x}$ the {\em malice-flipping distance}, and
$r_{\bf x}$ the {\em reject-flipping  distance}. We further define by $g_{\bf x}$ the gap between $m_{\bf x}$ and $r_{\bf x}$: $$g_{\bf x} = r_{\bf x} - m_{\bf x}.$$

The value of $g_{\bf x}$ is of particular interest.  If $g_{\bf x} < 0$, then the reject-flipping sample is still malicious, and therefore is an evading sample. 

It typically requires special crafting to embed an exploit in a malicious sample. Thus, it is expected that once a sample has turned to being benign, it is unlikely to regain the malicious functionality via random morphing. Similar assumption can be made about samples flipping from being accepted to rejected, for the features that the classifiers are trained to identify as grounds for making classification decision are unlikely to be formed via random morphing. Thus, we make an assumption that once a sample has flipped from being malicious (or rejected) to benign (or accepted), it would not be morphed to malicious (or rejected) again.

\subsection{Challenges in Evasion in the Dark}

There are two  main challenges in evasion with the limited blackbox accesses.  Firstly, the detector $\cal D$ only provides a binary output of accept/reject.  Hence, from  a set of  malicious samples, it is not clear which is ``closer''  to an evading sample.  Previous works such as \cite{evademl} assume that $\cal D$ returns a real-value confidence level, based on which the evasion is guided.  Nevertheless,  in practice, a more cautious detector might not reveal any information other than its final decision.

Secondly,  the blackbox morpher $\cal M$  alone does not provide flexibility in manipulating the samples. To illustrate, consider the situation where the adversary has a malicious sample $x_0$, and another sample  $x_1$ that losses its malicious functionality and being accepted by the detector. Intuitively, one could construct  another sample  $\hat{x}$  from $x_1$ and $x_0$ that has  higher chance of evasion.  With only blackbox accesses to $\cal M$, it is challenging to construct such $\hat{x}$. One possible way is to find a  sample  that lays somewhere in-between a path that starts from $x_1$ and ends at $x_0$. However,  it is not clear how to efficiently find a morphing path that  connects the two given samples. Even if it is possible to connect them, there is still a fundamental question on whether samples along the path have higher chance of evasion. 

\section{Proposed Evasion Methods}
\label{sec:proposed_method}

In this section, we present evasion methods that search for evading samples given a malicious sample that is rejected by the detector. For clarity of exposition, we first start with a trivial exhaustive  search algorithm \exhaust~which serves as the baseline for evaluation. We then optimize \exhaust~to reduce its operation costs (i.e., number of queries issued to $\cal D$ and $\cal T$), obtaining \evader. Next, we  propose a  hill-climbing based algorithm ${\tt EvadeHC}_0$ and finally present our main evasion approach  {\tt EvadeHC} that  further reduces the number of detector's queries.

\subsection{Evasion by Exhaustive  Search \exhaust}
\label{subsec:baseline}
On a malicious sample $x_0$ which is rejected by $\cal D$, \exhaust~operates by successively morphing $x_0$ and checking the morphed samples against $\cal D$ and $\cal T$ until an evading sample is found. \exhaust~is parameterised by a single threshold $L$ and proceeds as follows:

\begin{enumerate}
\item Choose a sequence of $L$ random seeds $ {\bf s} = \langle  s_1,s_2, \ldots,  s_L\rangle$. 
\item For each seed $s_i$ in $ {\bf s}$, obtain a morphed sample $x_i$ by invoking ${\cal M}(x_{i-1}, s_i)$. Check $x_i$ against $\cal D$ and $\cal T$, if it is evading, output $x_i$ and halts. Otherwise continue until $\bf s$ is consumed.
\item If  ${\bf s}$ has been consumed and an evading sample has not been found, repeat step 1 to 3 until an evading sample is found.

\end{enumerate}
Let $N_r$ be the number of iterations \exhaust~executed when the search halts. It is easy to see that the numbers of queries issued to $\cal D$, $\cal T$ and $\cal M$ are the same:
$$ N_d = N_t = N_m = N_r \cdot L $$

\subsection{Evasion by Binary Search \evader}

The baseline approach \exhaust~can be improved in various ways. One straight-forward enhancement is to reduce the number of detector queries issued in each path. In particular, one does not need to invoke $\cal D$ after every morphing step, but rather submit only the sample preceding the malice-flipping sample (e.g., $x_3$ in Figure 1a and $x_4$ in Figure 1b) to the detector. Further, one can also reduce the number of tester queries in searching for the malice-flipping samples by employing binary search in place of a series of sequential check as in the baseline.

The algorithm \evader, on an input $x_0$ which is malicious and rejected by $\cal D$, carries out the following: 
 \begin{enumerate}

 \item  Choose a sequence of $L$ random seeds ${\bf s}$, and invoke $\cal M$ successively with $s$ (starting from $x_0$) to obtain a path ${\bf x}$ of length $L$.
 
 \item  Find the malice-flipping sample $\tilde{x}$ in the path using binary-search with the tester ${\cal T}$. Let $\hat{x}$ be the immediate predecessor of $\tilde{x}$ on the path (i.e., $\hat{x}$ is malicious).
 
 \item  If $\hat{x}$ is accepted by the detector ${\cal D}$, output $\hat{x}$ and halt;  otherwise repeat step 1 to 3.
 \end{enumerate}

\indent Similar to \exhaust, \evader~is also parameterised by a single threshold $L$. The parameter $L$ is set to be sufficiently large such that with high probability, $L$  morphing steps would turn $x_0$ from malicious to benign.  In our experiments, $L$ ranges from $50$ to $80$.

\paragraph{Number of Queries. \ \ }  
We note that the number of paths \evader~traverses in finding evading samples is the same as the number of iterations taken by \exhaust, hence they incur the same number of morphing steps. On the other hand, \evader~requires less queries to $\cal D$ and $\cal T$. In particular, determining the malice-flipping sample on each path (step 2) incurs at most $ \lceil \log_2 L \rceil$ queries to ${\cal T}$, and there is only one query to the detector ${\cal D}$ per each path (step 3).  In sum, we have:
$$N_d= N_r, \ \ \ \ N_t\leq N_r \lceil \log_2 L \rceil, \ \ \ \  N_m = N_r \cdot L  $$

Note that the determination of the flipping samples is designed to minimise the number of queries to $\cal D$. In the scenario where $N_t$ imposes a larger cost than $N_d$, the algorithm can be modified so that the binary searches are conducted on $\cal D$, finding reject-flipping samples. It then tests the candidates against $\cal T$, incurring only one query to $\cal T$ per each path. 

\paragraph{Robustness of Binary Search. \ \ }

The binary search significantly reduces $N_t$ compared to the linear scan. As discussed earlier, we make an implicit assumption that once a sample has become benign (or accepted), it is highly unlikely that $\cal M$ will morph it to malicious (or rejected) again. Nevertheless, if such event happens, the binary-search may return a wrong flipping sample. Fortunately, the effect of such failure  is confined to the path in question, and does not affect other paths.

We can further improve the efficiency of the evasion by incorporating information obtained from $\cal D$ and $\cal T$ in a heuristic which ``guides''  the evasion. To this end, we propose a hill-climbing algorithm, dubbed ${\tt EvadeHC}_0$.

\subsection{Evasion by Hill-Climbing ${\tt EvadeHC}_0$}
\label{sec:hillclimb}
 ${\tt EvadeHC}_0$ is parameterised by two pre-determined integers $q_1$, $q_2$, a pre-determined real-value {\em jump factor} $0<\Delta <1$ and a  scoring function. The algorithm proceeds in iterations, maintaining a set $S$ of $q_2$ candidates over the iterations (in the first iteration, $S$ contains the given malicious samples $x_0$). In each iteration, ${\tt EvadeHC}_0$ first generates $q_1$ random paths with starting points originating from $S$. If any of the paths contains an evading sample, the algorithm outputs it and halts. Otherwise, a new candidate, determined by the jump factor $\Delta$, is chosen from each path, and a real-value score is computed for each new candidate using the scoring function. The scoring function takes as inputs the flipping distances of the path. Finally, $q_2$ candidates with the highest score are retained for the next iteration.

Details of each iteration are as follow:
\begin{enumerate}
\item
   For each sample $x$ in $S$, generates $\lfloor {q_1}/|S| \rfloor$  random paths with $x$ as starting point, obtaining approximately $q_1$ paths.    Let  $P$ be the set of  generated paths.  Reset $S$ to empty set.
\item
   For each path ${\bf x} \in { P}$, performs the following:
   \begin{enumerate}
      \item Find the   malice-flipping distance $m_{\bf x}$  using binary search. 
      \item  Find the reject-flipping distance $r_{\bf x}$ using binary search.  If $m_{\bf x}>r_{\bf x}$, output the reject-flipping sample and halt (i.e., an evading sample is found). 
      \item  Let  $\widehat{x}$ be the sample along ${\bf x}$ at a distance of $\lfloor \Delta m_{\bf x} \rfloor$  morphing steps from the starting point.  Compute the score $s=  {\tt score} (m_{\bf x}, r_{\bf x})$, and associate the sample $\widehat{x}$ with the score $s$.  Insert $\widehat{x}$ into ${S}$.
     \end{enumerate}
\item    
   Keep $q_2$ samples with the largest scores in $S$, and discard the rest . 
 \end{enumerate}

\begin{figure}
\centering
\includegraphics[height=4.6cm]{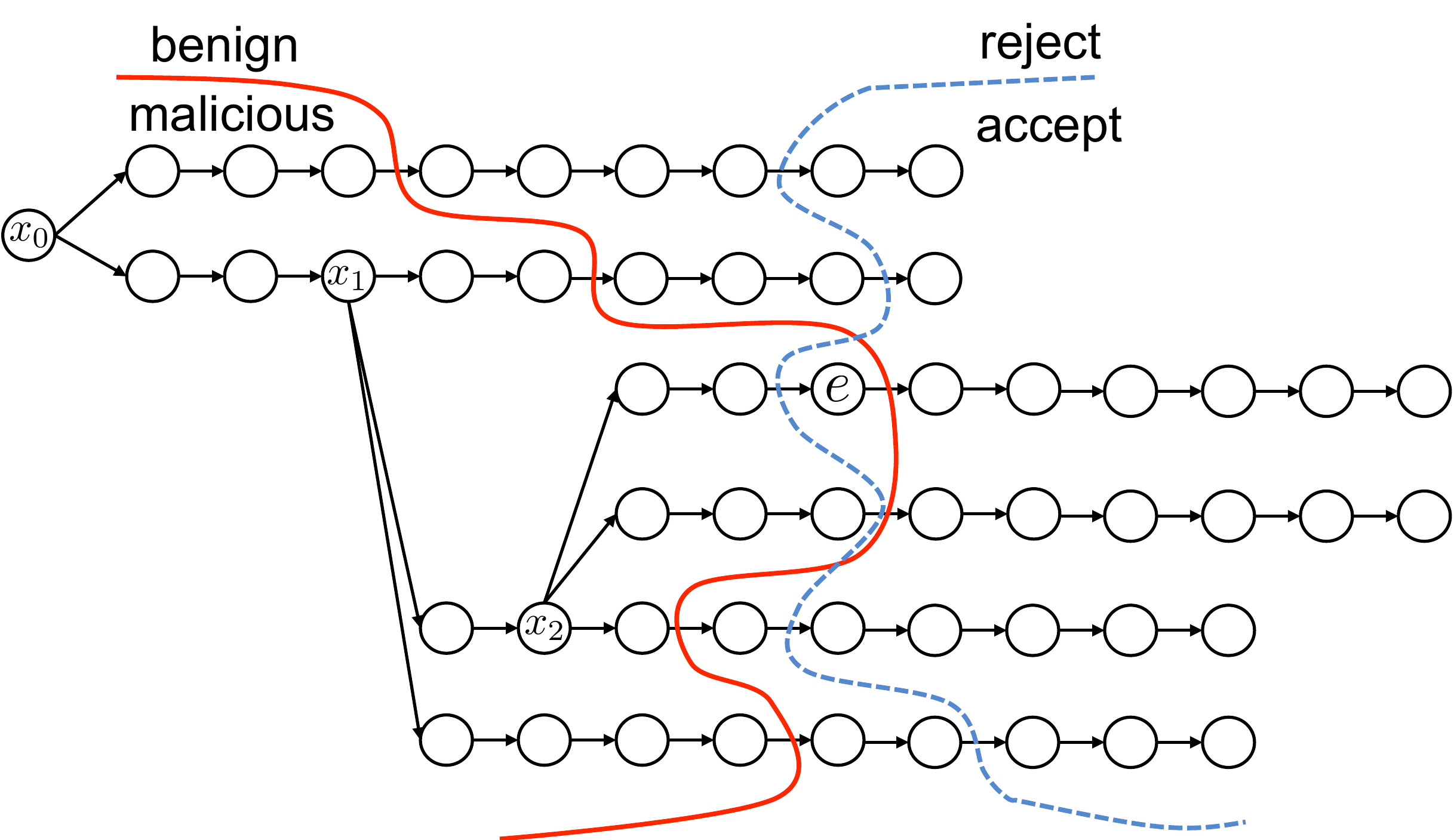} \\
\caption{Illustrating three iterations of Hill-Climbing.   The parameters $(q_1, q_2, \Delta)$ are $(1, 2, 0.5)$. Hence,  $S$ only contain one starting sample. The sample $x_0, x_1$ and $x_2$ are the starting sample chosen for the first, second and third iteration respectively. The sample $e$ is evading.   }
\label{fig:search}
\end{figure}

\paragraph{Scoring function.}
A crucial component is the scoring function {\tt score}.  Two possible choices of the scoring function are based on difference or relative ratio of the two flipping distances:    $$   {\tt score}_1 (m_{\bf x}, r_{\bf x}) = m_{\bf x} - r_{\bf x},   \ \ \  \  \ \   {\tt score}_2 (m_{\bf x}, r_{\bf x})= m_{\bf x}/r_{\bf x}.  $$

Next, we present an improved version of ${\tt EvadeHC}_0$, called \evadehc, that employs a branch-and-bound technique to further reduces the number of required detector queries.

\subsection{Enhanced Evasion by Hill-Climbing {\tt EvadeHC}} 
Recall that in ${\tt EvadeHC}_0$, a complete binary search is carried out on each path (Step 2(b)) to find a reject-flipping distance $r_{\bf x}$. Thus, it requires $q_1$ binary searches with the detector $\cal D$ to find the best $q_2$ candidates in each iteration. \evadehc, on the other hands, performs a {\em single} binary search to find a score $\hat{S}$ such that exactly $q_2$ candidates have scores greater than $\hat{S}$. 

Specifically, the binary search takes in  a set of paths $P$ as input. We assume that the malice-flipping  distance of each path in $P$ is already determined.  Let  $\hat{S}$ be the target score, and $\hat{P}$ be the target set of paths that attains scores greater than $\hat{S}$. \evadehc~searches for $\hat{S}$ and $\hat{P}$ in a single binary search, while  maintaining a set of candidate paths $C$ which is initially set to be $P$. 

We now describe a step in the binary search that  determines whether the target $\hat{S}$ is greater than a testing value  $S_0$. 
\begin{enumerate}
\item For each path ${\bf x}$ in  $C$, carry out the following:
	\begin{enumerate}
          \item Determine the smallest reject-flipping distance $r_{\bf x}$ such that the score is greater than $S_0$.  This does not require any query since the malice-flipping distance is known.  
           \item  Check whether this path attains a score larger than $S_0$.  This is achieved by querying $\cal D$, with the sample at distance $r_{\bf x}$ from the starting point.  If the sample is  accepted, then the score is greater than $S_0$.
    \end{enumerate}
\item If the total number of paths attaining score greater than $S_0$ is more than $q_2$, then $\hat{S}$ must be larger than $S_0$. Discard all paths whose score is lower than $S_0$, and recursively search on the upper half of $S_0$.
\item Otherwise,  $\hat{S}$ must be smaller than $S_0$, and all paths with score higher than $S_0$ must be in $\hat{P}$. Remove them from $C$ and recursively search on the lower half of $S_0$ for the remaining paths.
\end{enumerate}
 
 The applicability of this reduction depends on the choice of scoring function.  It is applicable as long as the scoring function is decreasing with respect to the reject-flipping distance $r_{\bf x}$,  which is the case for our choices of scoring functions ${\tt score}_1$ and ${\tt score}_2$.
 
 Similar to \evader,  the binary search is designed to  reduce $N_d$.  If queries to the tester  are more expensive than the  detector's queries,  we can swap  and apply the binary search  on the malice-flipping distances, and thus reducing $N_t$ instead.

\paragraph{Parameters $q_1, q_2, \Delta$. \ \ }  
Figure \ref{fig:search} illustrates the searching process. 
Intuitively,  a larger $q_1$ would improve accuracy of going toward the right direction while a larger $q_2$ would increase the robustness by recovering from local minimum, and a larger jump factor $\Delta$ could  reduce the number of iterations in good cases. Nevertheless, excessively large parameters might not increase the effectiveness of the search but rather incurs unnecessarily expensive performance cost. We examine in further details the effect of these parameters on the performance of \evadehc~in our experiments (Section~\ref{sec:eval}).

\section{Probabilistic Models}
\label{sec:model}
In this section, we explain the intuition behind the design of the proposed {\tt EvadeHC}. In general, for searching to make sense, each sample should have a  score-value for the  underlying ``state''.   We first give a way to assign a real-value state (which is also the score) to a sample  based on the binary  outcomes of $\cal T$ and $\cal D$ (``accept'' versus ``reject'' and ``malicious'' versus ``benign'').  Next, we propose a hidden state  model to capture the notion that the morphing process  is seemingly random. 

\subsection{States Representation} 
Given a sample $x$, we have 4 possible states from the binary outcomes of $\cal T$ and $\cal D$.   A searching strategy generally needs to select the ``best'' candidate from a given set of samples, and  this 4-state representation  alone would not provide  meaningful  information for selection.

Our main idea is to, ideally,  assign the probability  that a  random path  (generated by the morpher) starting from $x$  has a reject-flipping sample that is malicious\footnote{An alternative  choice is to consider the probability that    
a random path has an evading sample. However, this choice  is not suitable, since an arbitrary long path is likely to have an evading sample.}.

\paragraph{Evading Probability as the state.}
Given  a starting point $x$, let $M_{x}$ be the random variable of the malice-flipping distance on random path. Recall that the random paths arise from sequences of random seeds.   Likewise, let $R_{x}$ and $G_{x}$ be the random variable of the reject-flipping distance and the gap respectively.  It is not necessary that $M_{x} $ and $R_{x}$ are independent. Indeed, one would expect that they are highly positively correlated, since the detector attempts to detect malicious functionality. 
We shall revisit this later in Section~\ref{sec:eval}. At this points, readers can refer to Figure~\ref{fig:flipping_distances} for an intuition. The figure depicts flipping distances of $500$ morphed samples originating from the same malware (Pearson correlation coefficient is $0.34$).

For a sample $x$, let us assign the probability $\Pr ( G_{x} < 0)$ as its state. 
Recall that a path with negative gap implies that its reject-flipping sample is malicious and thus is an evading sample. In other words, the state of $x$ is the probability that a random path leads to an evading sample.    Now, suppose that the adversary has to pick one of two  candidates  $x$ and $y$ to continue the search,  then  the candidate with the larger state-value gives higher chance of finding an evading sample.   This comparison provides a way  for a hill-climbing algorithm to select candidates in each round. 

\paragraph{Expected flipping distances as the state.  } 
A main drawback  of explicitly taking  $\Pr (G_x <0)$ as  $x$'s state   is the high cost  in determining the probabilities during the search process.  Although one  may  estimate the distribution of $G_x$ by sampling multiple random paths, such accurate estimation would require extremely high number of queries, which in turn offsets the gain offered by efficient searching strategies.    Alternatively, we can take the expected reject-flipping and malice-flipping distances as the state.   Compare  to  the distribution $G_x$, the expected distances can be accurately estimated by drawing fewer random paths.  In  the proposed {\tt EvadeHC}, the measured flipping distances can be viewed as the estimates  of the expected distances.  From the expected distances, we can derive the expected gap.   Intuitively, the smaller the expected gap is, the higher the probability $\Pr (G_x <0)$ would be.  Although the expected flipping distances do not provide sufficient information for us to estimate the probability,  we can still employ the expected gap for comparison of probabilities.   This motivates the choices of the two scoring functions  ${\tt score}_1$  and ${\tt score}_2$ proposed in Section \ref{sec:hillclimb}.

\subsection{Hidden-State Random Morpher}
We consider an abstract model {\tt HsrMorpher}  for  analysis. Under {\tt HsrMorpher}, the morpher operates in the same way as the Random Oracle \cite{crypto_intro} employed in the studies of hash functions does, producing truly random outcomes for unique query, and being consistent  with repeated queries.  Based on {\tt HsrMorpher}, we give a condition whereby  \evadehc~performs poorly and cannot outperform \evader.

\paragraph{Randomly morphed sample.\ } 
Let us call a sample {\em seen}, if it has been queried against $\cal T, D$ or $\cal M$, or  has been an output of $\cal M$.  In other words, a sample is seen if it has been processed by  one of the blackboxes. A sample is {\em unseen} otherwise.    On a query consisting of a sample $x$ and a  random seed $s$, $\cal M$ first checks whether  the query $(x, s)$ has been issued before. If that is not the case,  $\cal M$ randomly and uniformly chooses an unseen sample to be the morphed sample\footnote{If the pool of unseen samples is empty, $\cal M$ simply halts and does not generate a morphed sample.}. Otherwise, the previously chosen sample will be outputted.  In additional, $\cal M$ also decides, in a random way, the hidden state  of the morphed sample, which is to be described next. 

\paragraph{Randomly reducing hidden values.\ }
Under {\tt HsrMorpher},  each sample $x$ is associated with a hidden state, represented by two real values $(a, b)$. There is a particular sample $x_o$ with state $(1,1)$ set to be seen initially, and it is also known to the adversary. Upon receiving a query $(x, s)$, $\cal M$ outputs a morphed sample $\widetilde{x}$ as described in the previous paragraph.   If  $x$ is unseen,  then the morpher sets  the state of $x$ to $(0, 0)$ and remembers that.   Next,  the morpher  selects two real values $(\alpha, \beta)$ from a random source $S$ and sets the hidden state of $\widetilde{x}$ as
$$   (a - \alpha,  b-\beta)$$
where $(a,b)$ is the hidden state of $x$.

The tester $\cal T$, when given a query $x$, decides the outcome based on the hidden state.  If $x$ is unseen, then $\cal T$ sets the hidden state to $(0, 0)$. Suppose the hidden state of $x$ is $(a,b)$, then $\cal T$ outputs {\em malicious}   iff  $a>0$.   Likewise, for the detector $\cal D$, it outputs {\em reject} iff $b>0$, and sets the state of $x$ to be $(0, 0)$ if it is unseen. 

Note that the only parameter in the {\tt HsrMorpher} model is the  random source $S$.

\subsection{Analysis of {\tt EvadeHC} on {\tt HsrMorpher} }
{\tt HsrMorpher} behaves randomly, and is arguably the worst situation the adversary can be in, as the adversary is unable to exploit domain knowledge to manipulating the samples.     For instance,  given two samples $x_0$ and $x_1$, it is not clear how to find a sequence of random  seeds that morph $x_0$ to $x_1$.

\paragraph{Ineffectiveness of \evadehc} To understand the performance of {\tt EvadeHC}, let us give a particular source $S$  for  {\tt HsrMorpher} that renders {\tt EvadeHC} ineffective. Let us consider a source $S$ where $\alpha$ and $\beta$ are discrete and only take on value 0 or 1.  Thus, all  possible hidden states are $(a, b)$ where $a, b$ are integers not greater than 1.   We can view the samples seen during an execution of {\tt EvadeHC} as a tree.  It is not difficult to show that the conditional probability $\Pr (G_x >0\  |\   \mbox{\em state of } x \ {\mbox{\em is }} (1,1) )$ is the same for all $x$ in this tree.   Thus,  in every iteration, all candidates have the same probabilities and it is irrelevant which is chosen for the next iteration. 

\paragraph{Random Source $S$. \ }  The model   {\tt HsrMorpher}  assumes that the reduction of the hidden values is independent of the  sample $x$.   To validate that real-life classifiers exhibit such property, in our empirical studies, we treat their internal classification score assigned to a sample as the sample's hidden state.   The empirical results demonstrate that  the distributions of the reduction are similar over a few selected samples.  

\paragraph{Remarks. \ }  One can visualise the search for evading sample  as a  race  in reducing the  hidden state $(a, b)$. Statistically, $a$ is expected to drop faster than $b$.  The evasion's goal is to find a random path that goes against all odds with  $b$ reduces to 0 before $a$ does.   Essentially, {\tt EvadeHC} achieves this in the following way:  It generates and instantiates a few  random paths.  The path with the smallest gap  arguably has $b$ reducing faster than average.  Hence,  {\tt EvadeHC} chooses a  point along this path as the starting point for the next iteration.  Figure \ref{fig:evasive_trace} depicts this process based on an actual trace obtained in our experimentation.

\section{Evaluation}
\label{sec:eval}

To study the feasibility and effectiveness of our proposed approaches, we conduct experimental evaluation on two well-known PDF malware classifier, namely \pdfrate~\cite{pdfrate} and Hidost~\cite{hidost}. We defer overview of PDF malwares and the two classifiers to Appendix~\ref{apd:case_studies}. We first describe our experimental setups and then reports the results.

\subsection{Experimental Setups}
\noindent\textbf{Morpher $\cal M$.}
We employ a simple morphing mechanism $\cal M$ that performs three basic operations with either insert, delete or replace an object in the original PDF file.  In order to perform these operations, the morpher $\cal M$ needs to be able to parse the PDF file into a tree-like structure, to perform the insertion, deletion or replacement on such a structure and finally to repack the modified structure into a variant PDF\footnote{To avoid redundant parsing of the same PDF file in multiple steps, $\cal M$ caches the modified tree structure (or the original tree structure of the malicious PDF file), and directly modifies it without having to parse the PDF file in each step.}. We employ a modified version of pdfrw~\cite{pdfrw} available at~\cite{pdfrw_modified} for parsing and repacking.

For insertion and replacement, the external objects that are placed into the original file is drawn from a benign PDF file. We collect ten benign PDF files of size less than 1MB via a Google search to feed into $\cal M$ as a source of benign objects. They are all accepted by the targeted detectors and confirmed by the tester $\cal T$ as not bearing any malicious behaviours. \\

\noindent\textbf{Tester $\cal T$.} 
The tester $\cal T$ is implemented using a malware analysis system called Cuckoo sandbox~\cite{Cuckoo}.  The sandbox opens a submitted PDF file in a virtual machine and captures its behaviours, especially the network APIs and their parameters. The PDF file is considered malicious if its behaviours involve particular HTTP URL requests and APIs traces. We follow previous work by Xu et al.~\cite{evademl} in selecting the HTTP URL requests and APIs traces as reliable signature in detecting malicious behaviours of the submitted PDF files. \\

\noindent\textbf{Dataset.}
We conduct the experimental studies using a dataset which consists of $500$ malicious PDF files that had been selected by Xu et al. in their experiments~\cite{evademl}. These samples are drawn from the Contagio~\cite{contagio} dataset, satisfying the following three conditions. First, they exhibit malicious behaviours observed by the tester. Secondly, they have to be compatible with the morphing mechanism (i.e. can be correctly repacked by pdfrw). And lastly, they are all rejected by both targeted detectors. We shall refer to these as malware seeds hereafter.  \\

\begin{figure}[t]
\centering
\begin{subfigure}{0.22\textwidth}
\begin{tikzpicture}[scale = 0.45]
\begin{axis}[
ymin = 0,
ticklabel style = {font=\Huge},
x label style={font = \Huge},           
xlabel = Average No. of iterations,
]
\addplot [pattern = crosshatch dots,pattern color = cornellred,
    hist={
        bins=17,
        data min=2,
        data max=19
    }   
] table [y index=0] {eval_data/pdfrate_iterations.dat};
\end{axis}
\end{tikzpicture}
\caption{\pdfrate}
\label{subfig:pdfrate_iteration_hist}
\end{subfigure}
\begin{subfigure}{0.22\textwidth}
\begin{tikzpicture}[scale = 0.45]
\begin{axis}[
ymin = 0,
ticklabel style = {font=\Huge},
x label style={font = \Huge},           
xlabel = Average No. of iterations,         
]
\addplot [pattern = crosshatch dots,pattern color = blue,
    hist={
        bins=9,
        data min=4,
        data max=13
    }   
] table [y index=0] {eval_data/hidost_iterations.dat};
\end{axis}
\end{tikzpicture}
\caption{Hidost}
\label{subfig:hidost_iteration_hist}
\end{subfigure}
\caption{Histogram of average numbers of iterations \evadehc~needs in finding evading samples against the two detectors.}
\label{fig:iterations_hist}
\vspace{-10pt}
\end{figure}
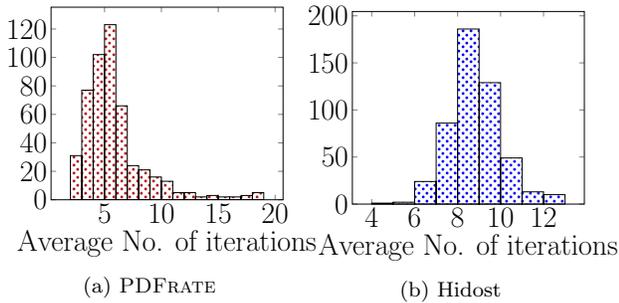

\noindent\textbf{Targeted detectors.} The targeted detectors in our experiments are based on the two PDF malware classifiers, namely \pdfrate~and Hidost. We make small changes to their original implementations~\cite{mimicus, hidost} such that the final outputs are binary decisions, rather than real-value scores. \\

\noindent\textbf{Scoring function.} The score function  adopted  in this studies is  ${\tt score}_1 (m, r)= m-r$, where  $m$ is  the malice-flipping distance and $r$  is the reject-flipping distance. \\

\noindent\textbf{Machine and other parameters.} All experiments are conducted on Ubuntu 14.04 commodity machines equipped with Intel(R) Xeon(R) CPU E5-2620 v3 $@$ 2.40GHz processors and  64GB of physical memory. The evasion process can be run in parallel, as there are no dependencies between different executions. In our experiments, we dedicate one machine to run $\cal D$ and $\cal M$, and nine others to run the tester $\cal T$. Each of these nine machines deploys $24$ virtual machines simultaneously, and is capable of performing approximately $2,000$ tester queries per hour.

We conducted five sets of experiments. The first set of experiments examines how the parameters $q_1$, $q_2$ and $\Delta$ effect the performance of \evadehc.  The second and third experiment sets evaluate the effectiveness of the proposed approaches in evading \pdfrate~and Hidost detectors, respectively. The efficiency of the evasion is determined by number of blackboxes queries and the overall running time. In the forth set of experiments, we consider a hypothetical setting where the detectors are hardened to make evasion more difficult and benchmark our approaches against the closely related work by Xu et al.\cite{evademl}, which relies on classification scores output by the detectors to guide the evasion. We emulate the hardening of detectors by reducing their defaults thresholds:   below $0.5$ for \pdfrate~and below $0$ for Hidost. The last set of experiments validates the {\tt HsrMorpher} model we discussed earlier in Section~\ref{sec:model}. Unless otherwise stated, the reported results are average values obtained over $10$ instances.

\subsection{Effect of Parameter Settings on \evadehc}
\begin{figure}[t]
\centering
\begin{subfigure}{.23\textwidth}
\begin{tikzpicture}[thick, scale = 0.45]
\begin{axis}[
	ticklabel style = {font=\Huge},
    xtick=data,
    xticklabels={$10$,$20$,$40$, $60$},
    ymin = 0,
    xlabel={$q_1$},
    ylabel={Average no. of iterations},
    x label style={font = \Huge},
    y label style={font = \Huge},
    legend style={at={(0.38,0.75)},anchor=west, draw=none,legend columns=1, font = \Huge},
]

\addplot [color = eggplant, mark=*,mark options={scale=1, solid, fill=eggplant}, densely dotted, thick] coordinates {
(1,38.5) 
(2,18.8) 
(3,6.83) 
(4,5.86) 
};

\addplot [color = deepcarrotorange, mark=square*,  mark options={scale=1,solid,fill=deepcarrotorange}, solid] coordinates { 
(1,19.53) 
(2,6.78) 
(3,6.34) 
(4,5.67) 

};

\addplot [color = charcoal, mark=diamond*,  mark options={scale=1,solid,fill=charcoal}, densely dashed] coordinates { 
(1, 23.7) 
(2, 7.02) 
(3,6.84) 
(4,5.87) 
};

\legend{$q_2=10\% \cdot q_1$,$q_2=25\% \cdot  q_1$,$q_2=50\% \cdot  q_1$}
\end{axis}
\end{tikzpicture}
\caption{Effect of $q_1$, $q_2$}
\label{subfig:effect_of_q1_q2}
\end{subfigure}
\begin{subfigure}{.23\textwidth}
\begin{tikzpicture}[thick, scale = 0.45]
\begin{axis}[
	ticklabel style = {font=\Huge},
	ybar,
    xtick=data,
    xticklabels={$0.4$, $0.5$, $0.6$,$0.7$,$0.8$, $0.9$},
    ymin = 0,
    enlarge x limits=0.16,
    bar width=20pt,  
    xlabel={$\Delta$},
    ylabel={Average no. of iterations},
    x label style={yshift=-5pt, font = \Huge},
    y label style={font = \Huge},
    legend style={at={(0.38,0.75)},anchor=west, draw=none,legend columns=1, font = \Huge},
]

\addplot [color = eggplant, fill = charcoal] coordinates {
(1,10.81)
(2,10.36)
(3,9.27) 
(4,6.78) 
(5,6.95) 
(6,8.73) 
};

\end{axis}
\end{tikzpicture}
\vspace{-2pt}
\caption{Effect of $\Delta$}
\label{subfig:effect_of_delta}
\end{subfigure}
\hfill
\begin{subfigure}{.32\textwidth}
\begin{tikzpicture}[thick, scale = 0.56]
\begin{axis}[
	ticklabel style = {font=\huge},
    ybar,
    enlarge x limits=0.16,
    xtick=data,
    xticklabels={$10$,$20$,$40$, $60$},
    ymin = 0,
    bar width=12pt,  
    area legend,
    xlabel={$q_1$},
    ylabel={$N_d$},
    x label style={font = \huge},
    y label style={font = \huge},
    legend style={at={(-0.16,1.06)},anchor=west, draw=none,legend columns=3, font = \LARGE},
]

\addplot [color = eggplant, pattern color = eggplant,pattern=dots] coordinates { 
(1,1012) 
(2,978) 
(3,713) 
(4,921) 
};

\addplot [color = deepcarrotorange, pattern color = deepcarrotorange,pattern=crosshatch] coordinates { 
(1,501) 
(2,352) 
(3,651) 
(4,893) 

};

\addplot [color = charcoal, pattern color = charcoal,pattern=north east lines] coordinates { 
(1, 618) 
(2, 369) 
(3,722) 
(4,908) 
};

\legend{$q_2=10\% \cdot q_1$,$q_2=25\% \cdot  q_1$,$q_2=50\% \cdot  q_1$}
\end{axis}
\end{tikzpicture}
\vspace{-5pt}
\caption{Average $N_d$ w.r.t different $q_1$, $q_2$}
\label{subfig:q1_q2_on_Nd}
\end{subfigure}
\vspace{-5pt}
\caption{Effect of $q_1$, $q_2$ and $\Delta$ on performance of \evadehc}
\vspace{-5pt}
\label{fig:effect_of_params}
\end{figure}
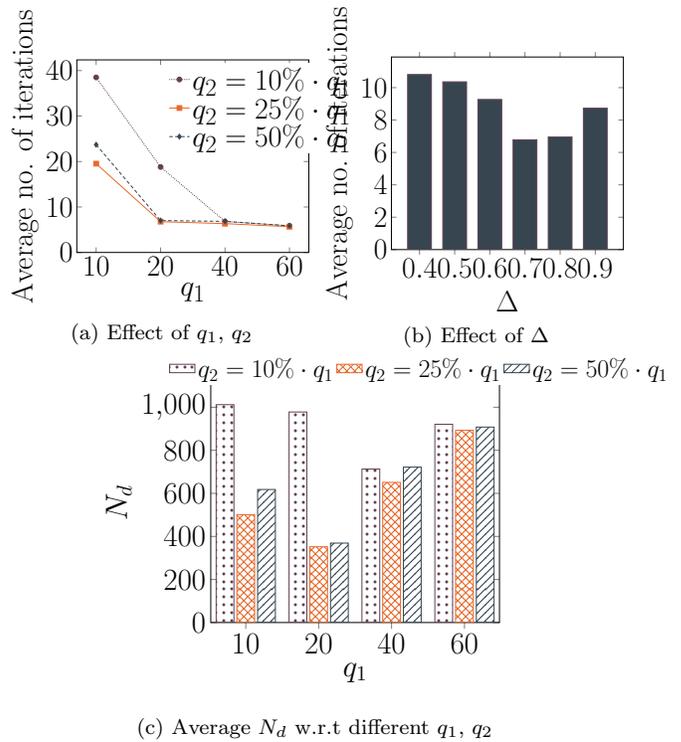
We first examine how the choice of parameters $q_1$, $q_2$ and $\Delta$ effects the performance of \evadehc. 
We run \evadehc~with different parameter settings on $100$ malware seeds randomly selected from our experimental dataset against \pdfrate~detector, varying $q_1$ from $10$ to $60$, and setting $q_2$ proportional to $q_1$ with a percentage ranging from $10\%$ to $50\%$. In addition, we vary the jump factor $\Delta$ from $0.4$ to $0.9$. The average numbers of iterations and amounts of detector queries required to find evading samples are reported in Figure~\ref{fig:effect_of_params}. 

Figure~\ref{subfig:effect_of_q1_q2} indicates that increasing $q_1$ will lessen the number of iterations. This is consistent with an intuition we discussed earlier, for larger $q_1$  would improve a likelihood that the algorithm is going toward the right direction. $q2$, on the other hands, has a more subtle effects on the number of iterations. In particular, a too small value may lead the search to a local minimum, hindering the search progress, while a too large value may accidentally bring ``bad'' candidates onto the next iterations. 
A similar trend is also observed on $\Delta$ (Figure~\ref{subfig:effect_of_delta}). When it is increased from $0.4$ to $0.7$, we witness a reduction in number of iterations. However, when it reaches $0.9$, the algorithm tends to loop through more iterations in order to find evading samples. The reason is that too small $\Delta$ limits progress the search can make in each iteration, while unnecessarily large $\Delta$ would increase the likelihood of samples in the next iterations being rejected.

It is worth noting that the number of detector queries required to find evading samples depends not only on the number of iterations, but also the value of $q_1$. Figure~\ref{subfig:q1_q2_on_Nd} shows average number of detector queries with respect to different settings of $q_1$ and $q_2$. 
While larger $q_1$ leads to smaller number of iterations, it may result in larger number of detector queries per each iterations, and thus larger queries overall (e.g., with $q_2 = 25\% \cdot q_1$, increasing $q_1$ from $20$ to $60$ leads to an increase in $N_d$). From the result of this experiment set, we choose $q_1 = 20$, $q_2 = 5$ and $\Delta = 0.75$ as the parameter setting for \evadehc~through all other sets of experiments.

\subsection{Evading \pdfrate~Detector}

The second experiment set focuses on evading \pdfrate~detector. All of our methods achieve $100\%$ evasion rates (i.e., found evading samples for all 500 malware seeds in our dataset). While the baseline \exhaust~and \evader~traverse $1048$ random paths on average to find an evading sample, \evadehc~only needs from three to six iterations (each iteration involves assessing $20$ random paths). Nevertheless, there are a few exceptions which require \evadehc~to take up to  $19$ iterations to find evading samples (Figure~\ref{subfig:pdfrate_iteration_hist}).

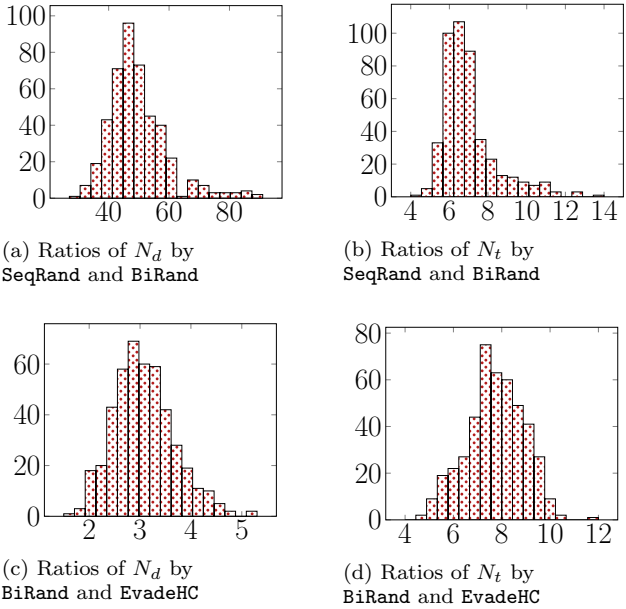
\begin{figure}[t]
\centering
\begin{subfigure}{.22\textwidth}
\begin{tikzpicture}[scale = 0.45]
\begin{axis}[
ticklabel style = {font=\Huge},
x label style={font = \Huge},           
y label style={font = \Huge},  
ymin = 0,      
]
\addplot [pattern = crosshatch dots,pattern color = cornellred,
    hist={
        bins=18,
        data min=27,
        data max=91
    }   
] table [y index=0] {eval_data/PDFRate_exhaust_randpath_detector_query_ratio.dat};
\end{axis}
\end{tikzpicture}
\caption{Ratios of $N_d$ by \\ \exhaust~and \evader}
\label{subfig:pdfrate_exhaust_evader_detector_compare}
\end{subfigure}
\hfill
\begin{subfigure}{.22\textwidth}
\begin{tikzpicture}[scale = 0.45]
\begin{axis}[
ticklabel style = {font=\Huge},
x label style={font = \Huge},           
y label style={font = \Huge},  
ymin = 0,      
]
\addplot [pattern = crosshatch dots,pattern color = cornellred,
    hist={
        bins=18,
        data min=4,
        data max=14
    }   
] table [y index=0] {eval_data/PDFRate_exhaust_randpath_oracle_query_ratio.dat};
\end{axis}
\end{tikzpicture}
\caption{Ratios of $N_t$ by \\ \exhaust~and \evader}
\label{subfig:pdfrate_exhaust_evader_oracle_compare}
\end{subfigure}
\begin{subfigure}{.22\textwidth}
\vspace{11pt}
\begin{tikzpicture}[scale = 0.45]
\hspace{3pt}
\begin{axis}[
ticklabel style = {font=\Huge},
x label style={font = \Huge},           
y label style={font = \Huge},  
ymin = 0,      
]
\addplot [pattern = crosshatch dots,pattern color = cornellred,
    hist={
        bins=18,
        data min=1.5,
        data max=5.3
    }   
] table [y index=0] {eval_data/pdfrate_detector_query_ratio.dat};
\end{axis}
\end{tikzpicture}
\caption{Ratios of $N_d$ by \\ \evader~and \evadehc}
\label{subfig:pdfrate_detector_compare}
\end{subfigure}
\hfill
\begin{subfigure}{.22\textwidth}
\vspace{11pt}
\begin{tikzpicture}[scale = 0.45]
\hspace{3pt}
\begin{axis}[
ticklabel style = {font=\Huge},
ymin = 0,
]
\addplot [pattern = crosshatch dots,pattern color = cornellred,
    hist={
        bins=18,
        data min=4,
        data max=12
    }   
] table [y index=0] {eval_data/pdfrate_oracle_query_ratio.dat};
\end{axis}
\end{tikzpicture}
\caption{Ratios of $N_t$ by \\ \evader~and \evadehc}
\label{subfig:pdfrate_oracle_compare}
\end{subfigure}
\caption{Performance comparison of \exhaust, \evader~and \evadehc~in evading \pdfrate.}
\label{fig:pdfrate_queries_no}
\end{figure}

We first compare the performance of \evader~to the baseline \exhaust. Figure~\ref{subfig:pdfrate_exhaust_evader_detector_compare} shows the ratios of their $N_d$, and Figure \ref{subfig:pdfrate_exhaust_evader_oracle_compare} plots that of $N_t$, clearly demonstrating the effectiveness of the binary search mechanism employed in \evader. Compared to the baseline, \evader~could reduce $N_d$ by upto $94$ times.

Next, we benchmark \evadehc~against \evader. Figure~\ref{subfig:pdfrate_detector_compare} report the ratios between their numbers of detector queries, and Figure \ref{subfig:pdfrate_oracle_compare} depicts ratios of their tester queries. \evader~typically demands from $2$ to $4$ times more detector queries, and $6$ to $10$ times more tester queries than \evadehc~does. More details in numbers of blackboxes queries required by \evadehc~and \evader~ are reported in Appendix~\ref{apd:pdfrate_performance}.

Compared to the baseline \exhaust, \evadehc~demands upto $450$ times fewer detector queries and $148$ times fewer tester queries, which translates to two orders of magnitude faster running times. We report running times of different approaches in Section~\ref{subsec:running_time}.

\begin{figure}[t]
\centering
\begin{subfigure}{.22\textwidth}
\begin{tikzpicture}[scale = 0.45]
\hspace{3pt}
\begin{axis}[
ticklabel style = {font=\Huge},
x label style={font = \Huge},           
y label style={font = \Huge},  
ymin = 0,      
]
\addplot [pattern = crosshatch dots,pattern color = blue,
    hist={
        bins=18,
        data min=50,
        data max=95
    }   
] table [y index=0] {eval_data/hidost_exhaust_randpath_detector_query_ratio.dat};
\end{axis}
\end{tikzpicture}
\caption{Ratios of $N_d$ by \\ \exhaust~and \evader}
\label{subfig:Hidost_exhaust_evader_detector_compare}
\end{subfigure}
\hfill
\begin{subfigure}{.22\textwidth}
\begin{tikzpicture}[scale = 0.45]
\hspace{3pt}
\begin{axis}[
ticklabel style = {font=\Huge},
x label style={font = \Huge},           
y label style={font = \Huge},  
ymin = 0,      
]
\addplot [pattern = crosshatch dots,pattern color = blue,
    hist={
        bins=18,
        data min=8,
        data max=12
    }   
] table [y index=0] {eval_data/hidost_exhaust_randpath_oracle_query_ratio.dat};
\end{axis}
\end{tikzpicture}
\caption{Ratios of $N_t$ by \\ \exhaust~and \evader}
\label{subfig:Hidost_exhaust_evade_oracle_compare}
\end{subfigure}
\begin{subfigure}{.22\textwidth}
\vspace{11pt}
\begin{tikzpicture}[scale = 0.45]
\begin{axis}[
ticklabel style = {font=\Huge},
ymin = 0,      
]
\addplot [pattern = crosshatch dots,pattern color = blue,
    hist={
        bins=18,
        data min=1.5,
        data max=4.5
    }   
] table [y index=0] {eval_data/hidost_detector_query_ratio.dat};
\end{axis}
\end{tikzpicture}
\caption{Ratios of $N_d$ by \\ \evader~and \evadehc}
\label{subfig:hidost_detector_compare}
\end{subfigure}
\hfill
\begin{subfigure}{.22\textwidth}
\vspace{11pt}
\begin{tikzpicture}[scale = 0.45]
\begin{axis}[
ticklabel style = {font=\Huge},
ymin = 0,
]
\addplot [pattern = crosshatch dots,pattern color = blue,
    hist={
        bins=18,
        data min=5,
        data max=22
    }   
] table [y index=0] {eval_data/hidost_oracle_query_ratio.dat};
\end{axis}
\end{tikzpicture}
\caption{Ratios of $N_t$ by \\ \evader~and \evadehc}
\label{subfig:hidost_oracle_compare}
\end{subfigure}
\caption{Performance comparison of \exhaust, \evader~and \evadehc~in evading Hidost.}
\label{fig:Hidost_queries_no}
\end{figure}
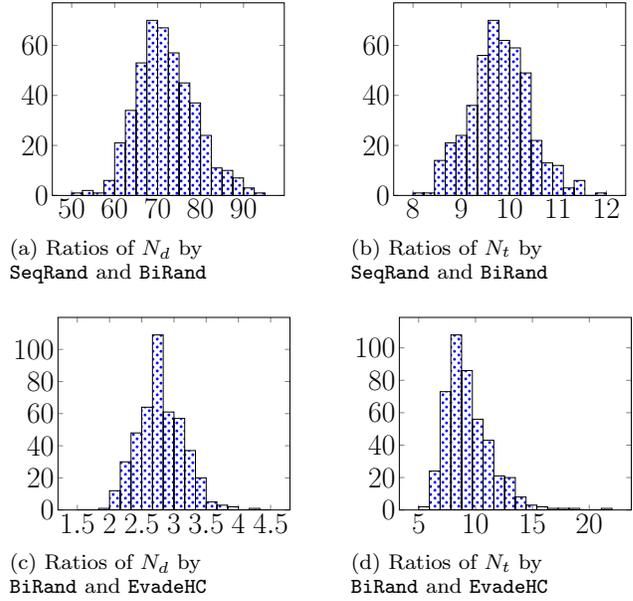

\subsection{Evading Hidost Detector}

In the third set of experiments, we evaluate our proposed approaches against Hidost detector. They also achieve $100\%$ evasion rate for the $500$ malware seeds in the dataset. For $95\%$ of the seeds, \evadehc~found the evading samples within $11$ iterations, while the other seeds need up to  $13$ iterations (Figure~\ref{subfig:hidost_iteration_hist}). \evader~and \exhaust, on the other hand, have to traverse approximately $1550$ paths on average to find an evading sample.

We report performance of \exhaust~in comparison with \evader~in Figure~\ref{subfig:Hidost_exhaust_evader_detector_compare} and \ref{subfig:Hidost_exhaust_evade_oracle_compare}, showing that \evader~is capable of outperforming \exhaust~by upto $92$ and $12$ times in term of detector and tester queries, respectively. 

Further, we also benchmark \evader~against \evadehc~by showing the ratios between their $N_d$ (Figures~\ref{subfig:hidost_detector_compare}) and $N_t$ (Figure \ref{subfig:hidost_oracle_compare}). The results show the superiority of \evadehc~over \evader. In particular, for most malware seeds, \evadehc~requires two to three times fewer detector queries and  seven to ten times fewer tester queries compared to \evader. More details in numbers of blackboxes queries required by
the two approaches are reported in Appendix \ref{apd:Hidost_performance}.

 When benchmarked against the baseline solution, \evadehc~attains upto two orders of magnitude lower execution cost, both in term of number of queries to the blackboxes and the overall running time.

\subsection{Execution Cost}
\label{subsec:running_time}

\begin{figure}[t]
\centering
\begin{tikzpicture}[thick, scale = 0.68]
\begin{axis}[
	ticklabel style = {font=\LARGE},
    ybar,
	ymode=log,   
    enlarge x limits=0.28,
    xtick=data,
    xticklabels={\exhaust, \evader, \evadehc},
    ymin = 0,
    bar width=17pt,  
    area legend,
    ylabel={Average Running Time (s)},
    x label style={yshift=-5pt, font = \LARGE},
    y label style={font = \LARGE},
    legend style={at={(0.62,0.85)},anchor=west, draw=none,legend columns=1},
]

\addplot [color = cornellred, pattern color = cornellred,pattern=grid] coordinates { 
(1,2574328/(9*25) 
(2,395543/(9*25) 
(3,53559/(9*25)
};

\addplot [color = blue, pattern color = blue,pattern=north west lines] coordinates { 
(1,5443124/(9*25) 
(2,553257/(9*25) 
(3,57924/(9*25) 
};

\legend{\pdfrate, Hidost}
\end{axis}
\vspace{-5pt}
\end{tikzpicture}
\caption{Average running time (in seconds) required to find an evading sample of different approaches. The results are taken over $5000$ evasions, $10$ for each of the malware seeds in our dataset}
\vspace{-10pt}
\label{subfig:running_time_compare}
\end{figure}
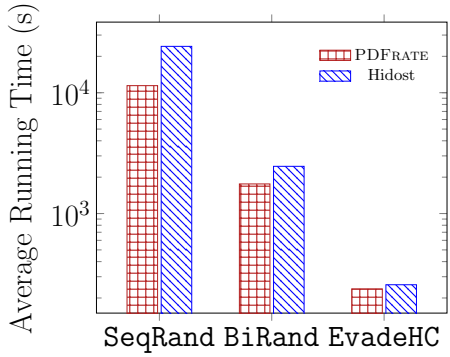

Figure \ref{subfig:running_time_compare} report average running time of different approaches in evading the two detectors. As discussed earlier, \evadehc~and \evader~ are order of magnitude more efficient than the baseline solution \exhaust. In particular, to find an evading sample against Hidost, \exhaust~takes on average $6.7$ hours, while \evader~and \evadehc~need only $40$ and $5$ minutes, respectively. 

It can also be seen that evading Hidost is often more expensive than evading \pdfrate. It would be interesting to investigate how the construct of Hidost provides resilience against the attacks.

We remark that \evader~and~\evadehc~are designed with a premise that $N_d$ is a dominating component in determining the evasion cost, thus minimising the number of detector queries, rather than tester queries which are computationally expensive. In situations where $N_t$ is the dominating component (e.g., applications for which computational cost is the main constrain), one can easily derive a variant of \evadehc~that applies the single  binary searches with $\cal D$ instead of $\cal T$. The performance of such algorithm is arguably similar to that of our current implementation, except for $N_d$ and $N_t$ whose values would be swapped. There is no impact on the accuracy, since the modification is purely on the algorithmic aspect. 

\subsection{Evading Trace Analysis}
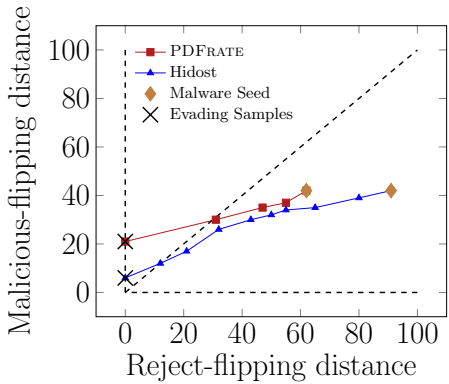
\begin{figure}
\centering
\begin{tikzpicture}[scale = 0.68]
\begin{axis}[
	xlabel=Reject-flipping distance,
  	ylabel=Malicious-flipping distance,
	legend style={legend columns=1, legend cell align=left,},
	legend style={draw = none,at={(0.589,0.958)}},
	ticklabel style = {font=\LARGE},
x label style={font = \LARGE},      
y label style={font = \LARGE},      
]
\addplot [color = cornellred, mark = square*] table { 
62	42
55	37
47	35   
31	30
0	21
};
\addplot [color = blue, mark=triangle*] table { 
91	42
80	39
65	35
55  34
50  32
43  30
32  26
21  17
12	12
0	6
};
\addplot [only marks,  mark = diamond*, mark options={scale=2, solid, color=brown}] table { 
91	42
62	42
};
\addplot [only marks,  mark = x, mark options={scale=3, thick, solid, color=black}] table { 
0	6
0	21
};
\addplot [dashed,mark=none, thick] table {
0	0
100	100
};
\addplot [thick, dashed,mark=none] table {
0	0	
0 	100
};
\addplot [thick, dashed,mark=none] table {
100	0	
0 	0
};

\legend{\pdfrate, Hidost, Malware Seed, Evading Samples}
\end{axis}
\end{tikzpicture}
\caption{Typical evading traces by \evadehc~against the two targeted detectors. A diagonal dashed line represents samples whose flipping distances are equal. Samples chosen in successive iterations usually move ``closer'' to the line.}
\label{fig:evasive_trace}
\end{figure}

To gain a better insight on how the hill-climbing heuristic works in \evadehc, we examine the flipping distances of samples generated along typical evading traces. An evading trace is a collection of samples that \evadehc~successively generates in finding an evading sample for a malware seed. Such a trace starts with an originally malicious PDF file, and ends with an evading sample. Samples generated in the first iteration would have reject-flipping distance larger than malice-flipping distance. \evadehc~would select promising samples from one iteration and continue to morph them in the next iteration so that they are eventually accepted by the detector before losing their malicious functionality. 

Figure~\ref{fig:evasive_trace} depicts typical evading traces that lead to evading samples against \pdfrate~and Hidost detectors.  The diagonal line in the figure represents points where malice-flipping distance and reject-flipping distance are equal, the vertical line represents points at which the detector's decision changes from reject to accept, and the horizontal line represents points where the malicious functionality of the samples are lost. Intuitively, the malware seeds' representations in term of malice-flipping distance and reject-flipping distance often lie below the diagonal line, and \evadehc~morphs them so that they move past the diagonal line, reaching the vertical line (i.e., being accepted) before crossing the horizontal line (i.e., losing the malicious functionality). It can also be seen from the figure that \pdfrate~is more evadable, as it is easier to move the samples generated against \pdfrate~past the diagonal compared to those that are evaluated against Hidost.

\subsection{Robustness against Hardened Detectors}
\label{subsec:evading_hardened_detectors}
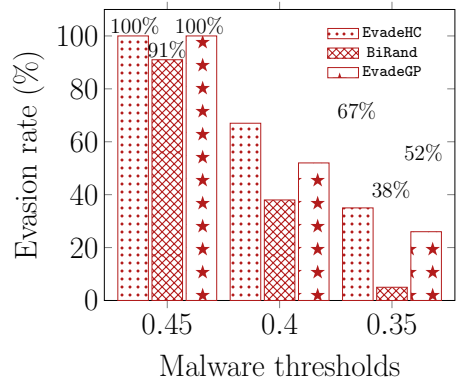
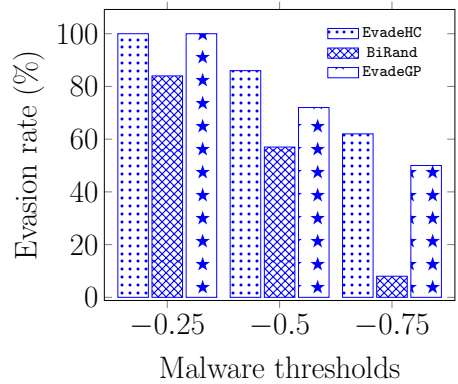
\begin{figure}[t]
\begin{subfigure}{0.45\textwidth}
\centering
\begin{tikzpicture}[thick, scale = 0.68]
\begin{axis}[
	ticklabel style = {font=\LARGE},
    ybar,
    enlarge x limits=0.28,
    xtick=data,
    xticklabels={$0.45$,$0.4$,$0.35$},
    ymin = 0,
    bar width=17pt,  
    area legend,
    xlabel={Malware thresholds},
    ylabel={Evasion rate ($\%$)},
    x label style={yshift=-5pt, font = \LARGE},
    y label style={font = \LARGE},
    legend style={at={(0.62,0.85)},anchor=west, draw=none,legend columns=1},
]

\addplot [color = cornellred, pattern color = cornellred,pattern=dots] coordinates { 
(1,100) 
(2,67) 
(3,35) 
};

\addplot [color = cornellred, pattern color = cornellred,pattern=crosshatch] coordinates { 
(1,91) 
(2,38) 
(3,5) 
};

\addplot [color = cornellred, pattern color = cornellred,pattern=fivepointed stars] coordinates { 
(1,100) 
(2,52) 
(3,26) 
};

\draw (1,104) node [xshift = -18pt] {\large $100\%$};
\draw (2,72) node [xshift = 43pt] {\large $67\%$};
\draw (3,39) node [xshift = 105pt] {\large $35\%$};

\draw (1,95) node [xshift = 0pt] {\large $91\%$};
\draw (2,42) node [xshift = 62pt] {\large $38\%$};
\draw (3,9) node [xshift = 124pt] {\large $5\%$};

\draw (1,104) node [xshift = 18pt] {\large $100\%$};
\draw (2,56) node [xshift = 80pt] {\large $52\%$};
\draw (3,30) node [xshift = 143pt] {\large $26\%$};

\legend{\evadehc,\evader, \evadegp}
\end{axis}
\end{tikzpicture}
\caption{Hardened \pdfrate}
\label{subfig:hardened_pdfrate}
\end{subfigure}%
\hspace{8pt}
\begin{subfigure}{0.45\textwidth}
\centering
\begin{tikzpicture}[thick, scale = 0.68]
\begin{axis}[
	ticklabel style = {font=\LARGE},
    ybar,
    enlarge x limits=0.28,
    xtick=data,
    xticklabels={$-0.25$,$-0.5$,$-0.75$},
    legend style={at={(0.3,0.85)},anchor=west, draw=none,legend columns=2},
    bar width=17pt,  
    area legend,
    xlabel={Malware thresholds},
    ylabel={Evasion rate ($\%$)},
    x label style={yshift=-5pt, font = \LARGE},
    y label style={font = \LARGE},
    legend style={at={(0.62,0.85)},anchor=west, draw=none,legend columns=1},
]

\addplot [color = blue, pattern color =  blue,pattern=dots] coordinates { 
(1,100) 
(2,86) 
(3,62) 
};

\addplot [color = blue, pattern color =  blue,pattern=crosshatch] coordinates { 
(1,84) 
(2,57) 
(3,8) 
};

\addplot [color = blue, pattern color =  blue,pattern=fivepointed stars] coordinates { 
(1,100) 
(2,72) 
(3,50) 
};

\draw (1,964) node [xshift = -18pt] {\large $100\%$};
\draw (2,826) node [xshift = 43pt] {\large $86\%$};
\draw (3,589) node [xshift = 105pt] {\large $62\%$};

\draw (1,795) node [xshift = 0pt] {\large $84\%$};
\draw (2,532) node [xshift = 62pt] {\large $57\%$};
\draw (3,35) node [xshift = 124pt] {\large $8\%$};

\draw (1,964) node [xshift = 18pt] {\large $100\%$};
\draw (2,680) node [xshift = 80pt] {\large $72\%$};
\draw (3,455) node [xshift = 143pt] {\large $50\%$};

\legend{\evadehc,\evader, \evadegp}
\end{axis}
\end{tikzpicture}
\caption{Hardened Hidost}
\label{subfig:hardened_Hidost}
\end{subfigure}%
\caption{Evasion rates of different approaches (displayed on top of the bars) against hardened detectors.}
\label{fig:evading_hardened_detector}
\end{figure}
In the forth experiment set, we investigate scenarios where the detectors are hardened to render evasions more difficult. We emulate hardened \pdfrate~by lowering its malware threshold down to $0.35$~\footnote{Authors of \pdfrate~\cite{pdfrate} reported that adjusting the malware threshold from $0.2$ to $0.8$ has negligible effect on accuracy.}, and hardened Hidost's to $-0.75$. Further, we bound the maximum number of detector queries that could be issued in finding an evading sample to $2,500$. If an evading sample can not found after the predefined number of detector queries, we treat the seed as non-evadable.

In addition, we benchmark the two approaches \evader~and \evadehc~against a technique by Xu et al.~\cite{evademl}, which we shall refer to as \evadegp. We stress that while \evadegp~relies on classification scores assigned to the samples to guide the search, our approaches do not assume such auxiliary information. We operate \evadegp~under a similar set of parameters reported in~\cite{evademl}: population size is $48$, mutation rate is $0.1$ and fitness stop threshold is $0.0$. We bound the number of generations that \evadegp~traverses to $50$ (instead of $20$ as in~\cite{evademl}), effectively limiting the number of detector queries incurred in each evasion to $2,500$. In addition, we disable its trace replay feature (i.e., mutation traces that successfully generates evading sample for one malware seed are not replayed on the other), treating each malware seed independently. This enables a fair basis to benchmark the effectiveness of our approaches against \evadegp, for ours do not assume trace replay.

Figure~\ref{fig:evading_hardened_detector} reports evasion rates of \evader~and \evadehc~in comparison with that of \evadegp~against the hardened detectors. For \pdfrate~detector with malware threshold set to $0.45$, both \evadehc~and \evadegp~attained $100\%$ evasion rate on our dataset, while \evader~only achieves $91\%$. When the malware threshold is further lowered to $0.4$ and $0.35$, the evasion rate of \evadehc~decreases to $68\%$ and $35\%$, respectively, while \evadegp~and \evader~witness more significant drops in their evasion rates, attaining evasion rate of as low as $5\%$ for \evader~and $26\%$ for \evadegp. Similarly, when the malware threshold of Hidost is reduced to $-0.75$, \evader~and \evadegp~could only find evading samples for $8\%$  and $50\%$ of the malware seeds respectively, while \evadehc~still attains an evasion rate of  $62\%$~(Figure~\ref{subfig:hardened_Hidost}). 

While it may come as a surprise that \evadehc~attains better evasion rate than \evadegp~even though it only assumes binary output instead of classification score as \evadegp~does, we suspect that this is because of the fact that \evadehc's scoring mechanism is capable of incorporating information obtained from both the tester and detector and thus is arguably more informative than the scoring mechanism of \evadegp~which mainly relies on the classification score output by the detector.

\subsection{Validating the Hidden-state Morpher Model}

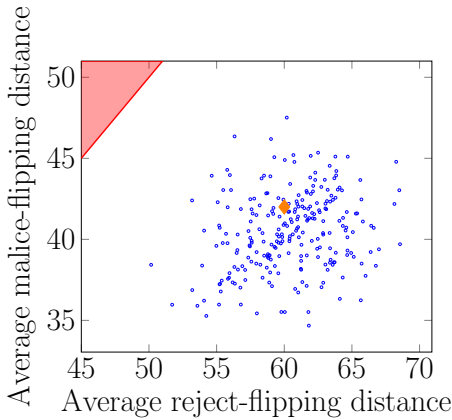
\begin{figure}[t]
\centering
\begin{tikzpicture}[thick, scale = 0.68]
\begin{axis}[
  xlabel=Average reject-flipping distance,
  ylabel=Average malice-flipping distance,
  	ticklabel style = {font=\LARGE},
x label style={font = \LARGE},      
y label style={font = \LARGE},   
xmin = 45,
ymax = 51, 
]
\addplot [only marks, mark=o, mark options={scale=0.38, solid, color=blue}] table {eval_data/flipping_distances.dat};
\addplot [thick,mark=none, color = red, fill=red, fill opacity = 0.38] table {
45 45
51 51
45 51
};

\addplot [only marks,  mark = diamond*, mark options={scale=2, solid, color=orange}] table { 
60	42
};

\draw (68,142) node {\color{red} Evading };
\draw (58,132) node {\color{red} Region};

\end{axis}
\end{tikzpicture}

\caption{Average flipping distances (with Hidost as the detector) of $200$ morphed samples originating from the same malware seed (depicted by the diamond). The malice-flipping distance of the chosen malware seed is $42$, and its reject-flipping distance is $60$. The pearson correlation coefficient of the distances is 0.34.}
\label{fig:flipping_distances}
\end{figure}
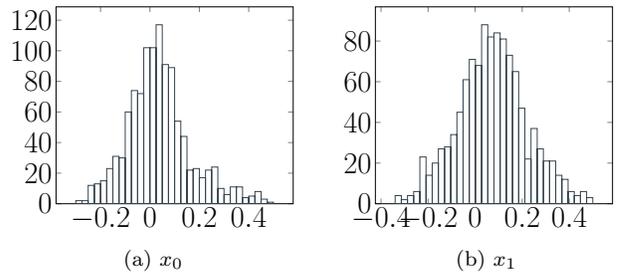
\begin{figure}[t]
\centering
\begin{subfigure}{.22\textwidth}
\begin{tikzpicture}[scale = 0.46]
\begin{axis}[
ymin = 0,
ticklabel style = {font=\Huge},
x label style={font = \Huge},           
]
\addplot [charcoal,
    hist={
        bins=32,
      	data min=-0.3,
 		data max=0.5
    }   
] table [y index=0] {eval_data/pdfrate_b2b_change2.dat};
\end{axis}
\end{tikzpicture}
\caption{$x_{0}$}
\label{subfig:score_change_x0}
\hfill
\end{subfigure}
\hfill
\begin{subfigure}{.22\textwidth}
\begin{tikzpicture}[scale = 0.46]
\begin{axis}[
ymin = 0,
ticklabel style = {font=\Huge},
x label style={font = \Huge},           
]
\addplot [charcoal,
    hist={
        bins=32,
      	data min=-0.34,
 		data max=0.5
    }   
] table [y index=0] {eval_data/pdfrate_b2b_change3.dat};
\end{axis}
\end{tikzpicture}
\caption{$x_{1}$}
\label{subfig:score_change_x1}
\hfill
\end{subfigure}
\vspace{-15pt}
\caption{Effect of morphing on \pdfrate~score. The histograms reported changes in classification scores after  $5$ morphing steps.}
\label{fig:score_change}
\end{figure}

To justify the proposition on states representation of a sample discussed earlier in Section~\ref{sec:model}, we  generate $200$ different morphed samples from the same malware seed, and measure their average flipping distances (the reject-flipping distance is measured against Hidost). The average flipping distances of each sample are computed from $20$ different random paths originating from it (the sequences of random seeds are the same for all samples), and plotted in Figure~\ref{fig:flipping_distances}. The figure also shows a certain correlation between the malice and reject flipping distances. Indeed, the pearson correlation coefficient of the distances is $0.34$, confirming our proposition earlier in Section~\ref{sec:model} that they are positively correlated.

 The proposed model   {\tt HsrMorpher}  assumes that the reduction of the hidden values is independent and identical for all the samples.   To validate that real-life classifiers exhibit such property,  we consider \pdfrate~and treat the internal classification score assigned to a sample as it's hidden state.   
We pick a sample $x_0$ and create another sample $x_1$ by applying a sequence of five morphing steps on $x_0$. We then generate $1,000$ random paths of length five from $x_0$. For each of the random path, we record the difference between classification scores of the last sample and $x_0$.  Figure \ref{subfig:score_change_x0}  shows the histogram of the recorded score differences. Similar experiment is carried out on $x_1$ and the histogram is shown in Figure \ref{subfig:score_change_x1}.  Observe that the two histograms are similar, giving evidences that  {\tt HsrMoprher} is appropriate.

We further study the validity of the {\tt HsrMorpher} model by running \evader~and \evadehc~under the abstract model. Recall that {\tt HsrMorpher} is parameterised by the random source $S$ that dictates how the two values $(\alpha, \beta)$ are chosen. From previous set of experiments, we observe that a typical malice-flipping distance is $42$, and reject-flipping distance is $60$ (Figure~\ref{fig:flipping_distances}). A more detailed analysis of the empirical data suggests that, with respect to the three blackboxes deployed in our experiments, $\alpha$ would follow normal distribution ${\cal N} (0.024, 0.01)$ and $\beta$ follows ${\cal N} (0.017, 0.011)$. The evasion starts with a malicious sample $x_o$ with state $(1,1)$ and succeeds if it can find an evading sample $e$ with state $(a, 0)$ for some $a > 0$. We repeat the experiment for $100$ times.

 Due to space constraint, we only show the average number of iterations (or paths) that the two approaches traverse in searching for evading samples under the {\tt HsrMorpher} model. The experimental results confirm 
the consistency of their behaviors under {\tt HsrMorpher} and the empirical studies. This strongly indicates that the abstract model can indeed serve as a basis to study and analyze the proposed evasion mechanism.

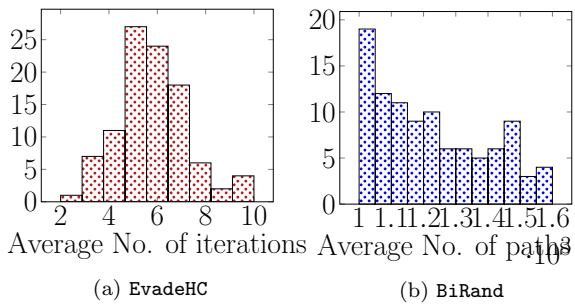
\begin{figure}[t]
\centering
\begin{subfigure}{0.22\textwidth}
\begin{tikzpicture}[scale = 0.45]
\begin{axis}[
ymin = 0,
ticklabel style = {font=\Huge},
x label style={font = \Huge},           
xlabel = Average No. of iterations,
]
\addplot [pattern = crosshatch dots,pattern color = cornellred,
    hist={
        bins=9,
        data min=2,
        data max=10
    }   
] table [y index=0] {eval_data/abstract_evadehc_iterations.dat};
\end{axis}
\end{tikzpicture}
\caption{\evadehc}
\label{subfig:abstract_evadehc_iterations}
\end{subfigure}
\begin{subfigure}{0.22\textwidth}
\begin{tikzpicture}[scale = 0.45]
\begin{axis}[
scaled x ticks= base 10:-3,
ymin = 0,
ticklabel style = {font=\Huge},
x label style={font = \Huge, xshift=-10pt},           
xlabel ={Average No. of paths},         
]
\addplot [pattern = crosshatch dots,pattern color = blue,
    hist={
        bins=12,
        data min=1000,
        data max=1600
    }   
] table [y index=0] {eval_data/abstract_evader_iterations.dat};
\end{axis}
\end{tikzpicture}
\caption{\evader}
\label{subfig:abstract_evader_paths}
\end{subfigure}
\caption{Histogram of average number of iterations and path that \evadehc~and \evader~needs in finding evading samples under the abstract model.}
\label{fig:validate_abstract_models}
\end{figure}

\section{Discussion}	
\label{sec:discussion}
In this section, we discuss potential mitigation strategies and implications suggested by our evasion attacks.

\subsection{Existing Defensive Mechanism}
Most existing works in evading learning-based classifier rely on real-value classification scores to guide the evasion~\cite{evademl, MLattacks_asiaccs17}. Thus, it has been suggested that hiding the classification scores or adding random noise to the scores may protect the classifier~\cite{hiding_score}. Another proposed defensive mechanism is to use a multi-classifier system whose classification score is randomly picked from different models trained with disjoint feature sets~\cite{random_picked_score}. Our results show that it is still feasible to evade learning-based classifiers without any classification score, rendering the above-mentioned defensive mechanisms ineffective.

\subsection{Potential Mitigation Strategies}
\textbf{Hardening the Malware Threshold.}
Previous work~\cite{evademl, pdfrate} suggested that the classification performance (i.e., false accept/false reject) is typically robust against the choice of the malware threshold, for original samples that are not adversarially manipulated  would have classification scores close to either of extremes. 
For example, the authors of \pdfrate~reported that adjusting the malware threshold from $0.2$ to $0.8$ has negligible effect on accuracy, because most samples would be assigned scores very close to either $0$ or $1$~\cite{pdfrate}. On the other hand, our experimental study (Section~\ref{subsec:evading_hardened_detectors}) suggests that even a slight change of the malware threshold could have significant impact on the evasion difficulties. Thus, it seems one potential mitigation strategies is to set the threshold to be more ``restrictive''. This is worth investigating in future works. \\

\noindent\textbf{Randomization.}
Another potential mitigation strategy is to embed into the classifiers a certain level of randomness. In particular, for samples whose classification scores are very close to the threshold, the classifiers can flip its classification decision with some probability. Given a proposition that most samples would have classification scores distant from the threshold, the above-mentioned flipping mechanism would not have significant impact on the classification accuracy. Similar to the previous mitigation strategy, more study is needed to conclude an effect of this countermeasure. \\

\noindent\textbf{Identifying Evading Attempts/Samples.}
In most cases, evading samples are found after hundreds of queries to the detector $\cal D$. In addition, the samples queried against $\cal D$ over the course of the evasion would resemble one another, to a certain extent. Thus, one possible method to detect evading attempts/samples is to have $\cal D$ remember samples that have been queried and apply some similarity/clustering-based techniques to identify evading queries/samples. In other words, $\cal D$ has to remain stateful.

\subsection{Evasion for Defense}
\label{subsec:adversarial_training}
Ironically, an effective evasion algorithm can also be used to build more secure classifiers that are robust against evasion attacks. Indeed, previous works have suggested that one can enhance the robustness of a learning based model by injecting adversarial examples throughout its training~\cite{adversarial_training, MLattacks_asiaccs17}. In particular, those morphed samples that retain the desired malicious activity, especially the ones that are misclassified by the learning based systems, can be included in the trainning dataset for the next step of training. We remark that by formulating the morphing process as random but repeatable, our approaches are capable of generating adversarial examples with great diversity, further enhancing effectiveness of the adversarial training.

\section{Related Work}
\label{sec:related_work}
Evasion attacks against classifiers have been studied in numerous contexts~\cite{evade_spam_detection, evade_adv_ad, hidost, evademl}. These works differ in their assumptions on the adversary's knowledge about the detector and how the data could be manipulated. Intuitively, detailed knowledge of the classifier significantly benefits the adversary in conducting evasion attacks. {\v{S}}rndic et al.~\cite{evade_adhoc_14} proposed a taxonomy of evasion scenarios based on the knowledge about the targeted classifier's components that are available to the adversary. These components include {\em training datasets} and {\em classification algorithms} together with their parameters, as well as a {\em feature set} that the classifier has learnt.

Various attacks against learning-based classifiers~\cite{evasion_test_time, evade_adhoc_14, face_reg_attack} assume that the adversary has high level of knowledge about the target system's internals (e.g., feature sets and classification algorithms), and could directly manipulate the malicious seeds in the feature space. It is unclear how these works could be extended to the constrained scenario we consider in this work, wherein the adversary does not have any knowledge of the feature set and could not manipulate the data on the feature space.

Xu et al.~\cite{evademl} relax an assumption on the high level of knowledge about the detector's feature set, only presume that the adversary is aware of the manipulation level of the morphing mechanism and has access to the classification scores output by the target detector. On the contrary, our technique does not assume any of such knowledge. Interestingly, we show in Section~\ref{subsec:evading_hardened_detectors} that even without relying on classification scores, our proposed algorithm \evadehc~still attains higher evasion rate against hardened detectors in comparison with previous work.

The evasion attacks proposed by Papernot et al.~\cite{MLattacks_asiaccs17} require training a local model that behaves somewhat similar to the targeted system, then search for evading samples against such local model, and show that they are also misclassified by the targeted system. The actual evasion happens on the substituting model whose internal parameters and training dataset are available to the adversary. Further, this approach heavily relies on the transferability of the adversarial samples. Our solutions, on the other hand, directly search for evading samples against the targeted classifier without necessitating training a local substituting model or making any assumption on the transferability of samples. 

A line of works on adversarial learning~\cite{model_stealing, membership_infer} also assume blackbox accesses to the targeted systems, but are different from ours in their adversarial goals. While Tram{\`e}r et al.~\cite{model_stealing} attempted to extract exact value of each model parameter and Shokri et al.~\cite{membership_infer} were interested in inferring if a data record was in the targeted model's training dataset, our focus is on deceiving the target system into misclassifying evading samples.

\section{Conclusion}
\label{sec:conclusion}
We have described~\evadehc, a generic hill-climbing base approach to evade binary-outcome detector using blackbox morphing, assuming minimal knowledge about both the detector and the data manipulation mechanism. We have demonstrated the effectiveness of the proposed approach against two PDF malware classifiers. Although the experiment studies are performed on malware classifier, the proposed technique and its security implications may also be of wider application to other learning-based systems. 
\section*{Acknowledgements}
We thank Amrit Kumar and Shiqi Shen for helpful discussions and feedback on the early version of the paper. This research is supported by the National Research Foundation, Prime Minister's Office, Singapore under its Corporate Laboratory$@$University Scheme, National University of Singapore, and Singapore Telecommunications Ltd. All opinions and findings expressed in this work are solely those of the authors and do not necessarily reflect the views of the sponsor.


\bibliographystyle{abbrv}
\bibliography{paper}

\appendix

\section{PDF Malware Classifiers}
\label{apd:case_studies}
In this section, we give an overview of the Portable Document Format (PDF) and PDF malwares, and briefly introduce the two detectors, namely \pdfrate~\cite{pdfrate} and Hidost~\cite{hidost}.

\subsection{PDF}
The Portable Document Format (PDF) is a standardised file format that is designed to decouple the document presentation from the underlying environment platform (e.g., application software, hardware or even OSes), thus enabling consistent presentation of the document across different platforms~\cite{pdf_ref}. A typical PDF file consists of four parts. The first part -- {\em header} -- contains  the magic number and the version of the format. The second part -- {\em body} -- incorporates a set of PDF objects comprising the content of the file. The third part -- {\em cross-reference table} (CRT) -- is an index table that gives the byte offset of the objects in the body. The last part -- {\em trailer} -- includes references to the CRT and other special objects such as the root subject. We depict an example of a PDF file in Figure~\ref{fig:pdf_example}.

The header, CRT and trailer are introduced with \textsf{\%PDF}, \textsf{xref} and \textsf{trailer} keywords, respectively. Objects in the body of the file may be either \textit{direct} (those that are embedded in another object) or \textit{indirect}. The indirect objects are numbered with a pair of integer identifiers and defined between two keywords \textsf{obj} and \textsf{endobj}. Objects have eight basic types which are Booleans (representing \textit{true} or \textit{false}), numbers, strings (containing 8-bit characters), names, arrays (ordered collections of objects), dictionaries (sets of objects indexed by names), streams (containing large amounts of data, which can be compressed) and the null objects. It is worth mentioning that there are some dictionaries associated with special meanings, such as those whose type is \textsf{JavaScript} (containing executable JavaScript code).

\begin{figure}
\centering
\resizebox{0.4\textwidth}{!}{
\begin{tabular}{|l|l|}
\hline
Header                   & \textsf{\%PDF}-1.5 \\ \hline                                                                          
\multirow{5}{*}{Body}    & 1 0 \textsf{obj} $<<$\textsf{/Count 1/Kids{[}7 0 R{]}/Type/Pages}      \\
                         & $>>$\textsf{endobj} \\
                         & $\cdots$ \\
                         & 21 0 \textsf{obj} null  \\
                         & \textsf{endobj} \\ \hline
\multirow{5}{*}{CRT}     & \textsf{xref} \\                                                                                              
                         & 0 22 \\                                                                                               
                         & 0000000002 65535 f \\                                                                                 
                         & $\cdots$ \\
                         & 0000000394 00000 n \\ \hline                                                                                
\multirow{5}{*}{Trailer} & \textsf{trailer} \\
                         & $\cdots$ \\
                         & \textsf{startxref} \\                                                                                      
                         & 1637 \\
                         & \textsf{\%\%EOF} \\                                                                                          \hline
\end{tabular}
}
\caption{An example of a PDF file structure}
\label{fig:pdf_example}
\end{figure}

\subsection{PDF Malwares}
Due to its popularity, PDF files have been extensively exploited to deliver malware. In addition, given an increasing number of vulnerabilities in Acrobat readers reported recently~\cite{CVE_Acrobat}, the threat of PDF malwares is clearly relevant. The malicious payploads embedded in the PDFs are often contained in JavaScript objects or other objects that could exploit vulnerabilities of a particular PDF reader in use.

\subsection{Detectors}

Various PDF malware detectors have been proposed in the literatures. A line of works~\cite{Wepawet,PJScan} targeted  JavaScript code embedded in the malicious PDFs (or PDF malware). They first extract a JavaScript code from the PDF, and either dynamically or statically analyse such a code to assess the maliciousness of the PDF. Nevertheless, these works would fail to detect PDF malware wherein the malicious payloads are not embedded in JavaScript code or the code itself is hidden~\cite{selvaraj_rise}. Hence, recent works have taken another approach, relying on the structural features of the PDFs, to perform static detection. Structural feature-based detectors perform detection based on an assumption that there exist differences between internal object structures of benign and malicious PDF files.

Our experimental evaluations are conducted against two state-of-the-art structural feature-based detectors, namely \pdfrate~\cite{pdfrate} and Hidost~\cite{hidost}. These two systems are reported to have very high detection rate on their testing datasets, and are also studied in previous works on evading detection~\cite{evademl}. It is worth mentioning that the outputs of these two detectors are real-value scores which are to be compared against thresholds to derive detection results. In our experiments, we make small changes to their original implementations so that they return binary outputs.

\paragraph{\pdfrate.}
\pdfrate~is an ensemble classifier consisting of a large number of classification trees. Each classification tree is trained using a random subset of the training data and based on an independent subset of features. These features include object keywords, PDF metadata such as author or creation data of the PDF file, and several properties of objects such as lengths or positions. At the time of classification, each tree outputs a binary decision indicating its prediction on the maliciousness of the input. The output of the ensemble classifier is the fraction of trees that consider the input ``malicious'' (a real-value score raning from $0$ to $1$). The default cutoff value is set at $0.5$. 

\pdfrate~was trained using $5,000$ benign and $5,000$ malicious PDF files randomly selected from the Contagio dataset~\cite{contagio}. The classifier consists of $1,000$ classification trees each of which covers $43$ features. The total number of features covered by all the classification trees is $202$ (but only $135$ are documented in \pdfrate's documentation). An open-source implementation of \pdfrate, namely Mimicus~\cite{mimicus}, is by {\v{S}}rndi{\'c} et al.~\cite{evade_adhoc_14}. We utilise this implementation in our experiments.

\paragraph{Hidost.} Hidost is a support vector machine (SVM) classifier~\cite{hidost}. SVM aims to fit a hyperplane (which can be expressed using a small number of support vectors) to training data in such a manner that data points of both classes are separated with the largest possible margin. Hidost works by first mapping a data point (representing a submitted PDF file) to an indefinite dimensional space using radial basis function, and reports the distance between the data point and the hyperplane as a measurement of its maliciousness. If the distance is positive, the PDF file is considered malicious; otherwise,  the file is flagged as benign.

Hidost was trained using  $5,000$ benign and $5,000$ malicious PDF files. Hidost operates based on $6,087$ classification features, which are structural paths of objects. These structural paths are selected from a set of $658,763$ PDF files (including both benign and malicious instances) based on their popularity (i.e., each of them appeared in at least $1,000$ files). We use the implementation made available by the author of Hidost in our experiments~\cite{hidost}.

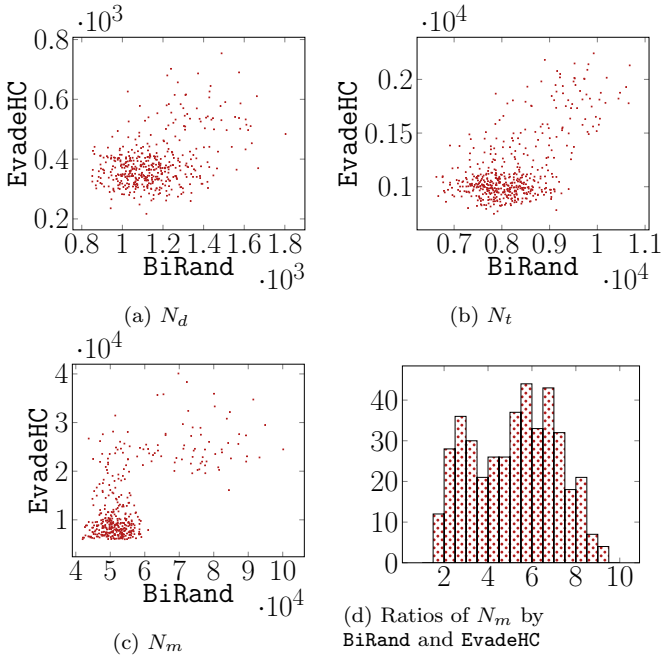
\begin{figure}[t]
\centering
\begin{subfigure}{.22\textwidth}
\centering
\begin{tikzpicture}[thick, scale = 0.45]
\hspace{-2pt}
\begin{axis}[
ticklabel style = {font=\Huge},
x label style={font = \Huge, yshift = 3pt},           
y label style={font = \Huge},           
ylabel= \evadehc,
xlabel= \evader,
scaled ticks=base 10:-3,
]
\addplot [only marks, mark=square*,mark options={scale=0.15, solid, color=cornellred}] table {eval_data/pdfrate_detector_queries.dat};
\end{axis}
\end{tikzpicture}
\caption{$N_d$}
\label{subfig:pdfrate_detector}
\end{subfigure}
\hspace{6pt}
\begin{subfigure}{.22\textwidth}
\begin{tikzpicture}[thick, scale = 0.45]
\begin{axis}[
ticklabel style = {font=\Huge},     
y label style={font = \Huge},    
x label style={font = \Huge, yshift = 3pt},             
ylabel= \evadehc,
xlabel= \evader,
scaled ticks=base 10:-4,
]
\addplot [only marks, mark=square*,mark options={scale=0.15, solid, color=cornellred}] table {eval_data/pdfrate_oracle_queries.dat};
\end{axis}
\end{tikzpicture}
\caption{$N_t$}
\label{subfig:pdfrate_oracle}
\end{subfigure}
\begin{subfigure}{.22\textwidth}
\begin{tikzpicture}[thick, scale = 0.45]
\hspace{9pt}
\begin{axis}[
 ticklabel style = {font=\Huge},
y label style={font = \Huge},           
ylabel= \evadehc,
xlabel= \evader,
scaled ticks=base 10:-4,
x label style={font = \Huge, yshift = 3pt},      
xmin = 39000,
ymax = 42000
]
\addplot [only marks, mark=square*,mark options={scale=0.15, solid, color=cornellred}] table {eval_data/pdfrate_morpher_queries.dat};
\end{axis}
\end{tikzpicture}
\caption{$N_m$}
\label{subfig:pdfrate_morpher}
\end{subfigure}
\hfill
\begin{subfigure}{.22\textwidth}
\vspace{12pt}
\begin{tikzpicture}[scale = 0.46]
\hspace{9pt}
\begin{axis}[
ticklabel style = {font=\Huge},
x label style={font = \Huge},           
y label style={font = \Huge},           
ymin = 0,
]
\addplot [pattern = crosshatch dots,pattern color = cornellred,
    hist={
        bins=18,
        data min=1,
        data max=10
    }   
] table [y index=0] {eval_data/pdfrate_morpher_query_ratio.dat};
\end{axis}
\end{tikzpicture}
\caption{Ratios of $N_m$ by \\ \evader~and \evadehc}
\label{subfig:pdfrate_morphing_compare}
\end{subfigure}
\caption{Average number of blackbox queries required in evading \pdfrate}
\label{fig:apd_pdfrate_performance}
\end{figure}
\begin{figure}[t]
\centering
\centering
\begin{subfigure}{.22\textwidth}
\begin{tikzpicture}[thick, scale = 0.45]
\hspace{-2pt}
\begin{axis}[
ticklabel style = {font=\Huge},
x label style={font = \Huge, yshift = 3pt},           
y label style={font = \Huge},           
ylabel= \evadehc,
xlabel= \evader,
scaled ticks=base 10:-3,
]
\addplot [only marks, mark=triangle*, mark options={scale=0.15, solid, color=blue}] table {eval_data/Hidost_detector_queries.dat};
\end{axis}
\end{tikzpicture}
\caption{$N_d$}
\label{subfig:Hidost_detector}
\end{subfigure}
\hspace{6pt}
\begin{subfigure}{.22\textwidth}
\vspace{0pt}
\begin{tikzpicture}[thick, scale = 0.45]
\begin{axis}[
ticklabel style = {font=\Huge},
x label style={font = \Huge, yshift = 3pt},           
y label style={font = \Huge},           
ylabel= \evadehc,
xlabel= \evader,
scaled ticks=base 10:-2,
]
\addplot [only marks, mark=triangle*, mark options={scale=0.15, solid, color=blue}] table {eval_data/Hidost_oracle_queries.dat};
\end{axis}
\end{tikzpicture}
\caption{$N_t$}
\label{subfig:Hidost_oracle}
\end{subfigure}
\begin{subfigure}{.22\textwidth}
\vspace{0pt}
\begin{tikzpicture}[thick, scale = 0.45]
\hspace{2pt}
\begin{axis}[
ticklabel style = {font=\Huge},
x label style={font = \Huge, yshift = 3pt},           
y label style={font = \Huge},           
ylabel= \evadehc,
xlabel= \evader,
scaled ticks=base 10:-4
]
\addplot [only marks, mark=triangle*, mark options={scale=0.15, solid, color=blue}] table {eval_data/Hidost_morpher_queries.dat};
\end{axis}
\end{tikzpicture}
\caption{$N_m$}
\label{subfig:Hidost_morpher}
\end{subfigure}
\hfill
\begin{subfigure}{.22\textwidth}
\vspace{11pt}
\begin{tikzpicture}[scale = 0.45]
\hspace{3pt}
\begin{axis}[
ticklabel style = {font=\Huge},
ymin = 0,
]
\addplot [pattern = crosshatch dots,pattern color = blue,
    hist={
        bins=15,
        data min=5,
        data max=20
    }   
] table [y index=0] {eval_data/hidost_morpher_query_ratio.dat};
\end{axis}
\end{tikzpicture}
\caption{Ratios of $N_m$ by \\ \evader~and \evadehc}
\label{subfig:hidost_morphing_compare}
\end{subfigure}
\caption{Average number of blackbox queries required in evading Hidost}
\label{fig:apd_hidost_performance}
\end{figure}
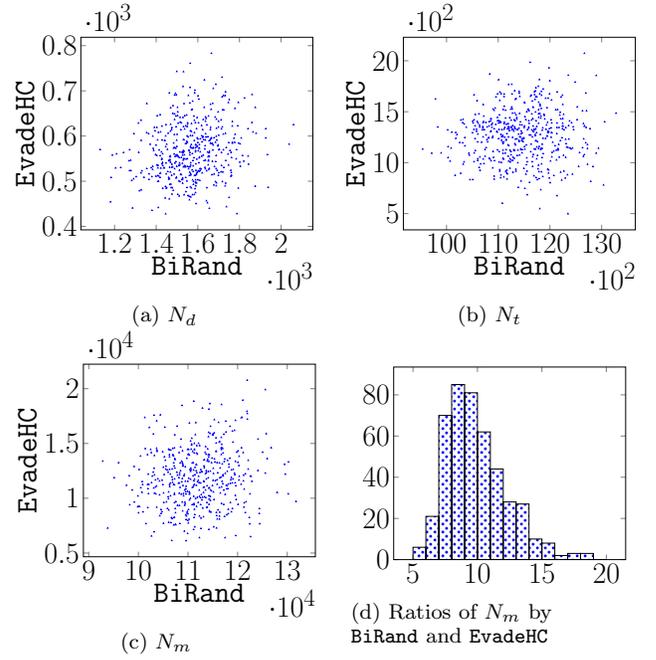

\section{Number of blackbox queries in evading \pdfrate}
\label{apd:pdfrate_performance}
The number of blackbox queries that \evader~and \evadehc~required in evading \pdfrate~are reported in Figures \ref{subfig:pdfrate_detector}, \ref{subfig:pdfrate_oracle}, \ref{subfig:pdfrate_morpher}. We do not report these metrics of the baseline, for they are all equal to the number of morphing steps that \evader~incurs.

As can be seen from Figure~\ref{subfig:pdfrate_detector}, for a majority of the malware seeds, \evadehc~needs at most $494$ queries, while \evader~requires approximately $1500$. The remaining seeds appear to be much more difficult to evade, for which \evadehc~and \evader~need up to  $786$ and $1837$ queries, respectively. Similarly, \evadehc~also requires fewer tester queries and less morphing efforts than \evader~(Figures \ref{subfig:pdfrate_oracle} and \ref{subfig:pdfrate_morpher}). In particular, for the majority of the malware seeds, \evadehc~can find an evading sample with less than $1,237$ tester queries and $21,707$ morphing steps, while \evader~consumes up to  $9,100$ and $63,288$ morphing steps. 

To get an insight why some seeds necessitated much more effort in finding evading samples, we check their classification scores given by \pdfrate. It comes as no surprise that their classification scores are higher than the rest of the malware seeds. In order words, it is harder to find evading samples for these seeds because the detector perceived their maliciousness more clearly.

The histogram of the ratios between the numbers of morphing steps required by the two approaches is depicted in Figure~\ref{subfig:pdfrate_morphing_compare}. \evadehc~requires as low as one tenth morphing efforts compared to \evader~in order to find an evading sample.

\section{Number of blackbox queries in evading Hidost}
\label{apd:Hidost_performance}
We report the amounts of blackbox queries \evadehc~and \evader~incur in evading Hidost in Figure ~\ref{subfig:Hidost_detector},~\ref{subfig:Hidost_oracle} and ~\ref{subfig:Hidost_morpher}. Overall, \evadehc~outperforms \evader~with respect to all three metrics $N_d$, $N_t$ and $N_m$. In particular, \evadehc's requires as few as $427$ detector queries, while \evader~needs as least $1,131$ queries to find an evading sample. With respect to the tester, \evadehc~requires no more than $2073$ queries, while \evader~requires $11,500$ queries on average. The similar trend can also be observed on morphing effort, wherein \evadehc~requires about $12,500$ morphing step on average, while such number for \evader~is approximately $10$ times larger (Figure~\ref{subfig:hidost_morphing_compare}).

\end{document}